\documentclass[useAMS,usenatbib]{mn2e}
\pdfoutput=1

\usepackage{graphicx}    % Include figure files
\usepackage{dcolumn}     % Align table columns on decimal point
\usepackage{bm}          % bold math
\usepackage{amssymb,amsmath}  
\usepackage{color}
\usepackage{fixltx2e}
\usepackage{hyperref}
\hypersetup{colorlinks=true,citecolor=blue} % to make the paper nicer

\title[Homogeneity and isotropy in 2MPZ]
{Homogeneity and isotropy in the 2MASS Photometric Redshift catalogue}
\author[D. Alonso, A. I. Salvador, F. J. S\'{a}nchez, M. Bilicki,
        J. Garc\'{i}a-Bellido, E. S\'{a}nchez]
{D. Alonso$^1$\thanks{E-mail: david.alonso@astro.ox.ac.uk}, A. I. Salvador$^2$,
 F. J. S\'{a}nchez$^3$, M. Bilicki$^{4,5}$,\newauthor
J. Garc\'{i}a-Bellido$^2$, E. S\'{a}nchez$^3$\\
 $^{1}$University of Oxford, Denys Wilkinson Building, Keble Road, Oxford, OX1 3RH,  UK\\
 $^{2}$Instituto de F\'{i}sica Te\'{o}rica, UAM-CSIC, Madrid, Spain.\\
 $^{3}$Centro de Investigaciones Energ\'{e}ticas Medioambientales y Tecnol\'{o}gicas,
       CIEMAT, Madrid, Spain.\\
 $^{4}$Astrophysics, Cosmology and Gravity Centre (ACGC), Department of Astronomy,
       University of Cape Town, South Africa.\\
 $^{5}$Kepler Institute of Astronomy, University of Zielona G\'{o}ra, Poland.
}
\begin{document}
  \date{\today}
  \pagerange{1--18} \pubyear{2015}
  \maketitle

\begin{abstract}
  Using the 2MASS Photometric Redshift catalogue we perform a number of statistical tests aimed at
  detecting possible departures from statistical homogeneity and isotropy in the large-scale
  structure of the Universe. Making use of the angular homogeneity index, an observable proposed
  in a previous publication, as well as studying the scaling of the angular clustering and number
  counts with magnitude limit, we place constraints on the fractal nature of the galaxy
  distribution. We find that the statistical properties of our sample are in excellent agreement
  with the standard cosmological model, and that it reaches the homogeneous regime significantly
  faster than a class of fractal models with dimensions $D<2.75$. As part of our search for 
  systematic effects, we also study the presence of hemispherical asymmetries in our data, finding 
  no significant deviation beyond those allowed by the concordance model.
\end{abstract}


\section{Introduction}\label{sec:intro}
  In the last decades, due to the increasing abundance and quality of astronomical observations,
  we have been able to draw a fairly complete picture of the Universe on cosmological scales. The
  so-called $\Lambda$CDM ($\Lambda$  - Cold Dark Matter) model can successfully explain the vast
  majority of observational data, and we are now able to constrain the value of many of
  its free parameters to percent precision \citep{2014A&A...571A..16P}. This model is based on
  a small number of premises, arguably the most fundamental of which is the Cosmological
  Principle (CP), which states that on large scales the distribution of matter in the Universe is
  homogeneous and isotropic.
  
  The exact validity of the CP is unfortunately difficult to verify. While the high degree of
  isotropy of the Cosmic Microwave Background (CMB) \citep{1996ApJ...473..576F,2014A&A...571A..23P}
  certainly supports this assumption in the early Universe, as well as during most of its history,
  it is not possible to unequivocally relate that to the degree of homogeneity of the present-day
  Universe\footnote{It is worth noting that the late-time, non-linear evolution of density
  perturbations can potentially affect the background expansion of the Universe in a process
  called ``back-reaction'' \citep{2011CQGra..28p4008R}. However, since general
  back-reaction models preserve statistical homogeneity and isotropy, they cannot be
  constrained by our analysis.}. Large-scale homogeneity and isotropy are usually taken for granted
  without proof in the application of many cosmological probes \citep{2011RSPTA.369.5102D}. This
  is often a reasonable approach, as long as the assumptions that go into the analysis methods
  are clearly stated and understood. However, since multiple 
  cosmological observables rely on the validity of the CP, it would be desirable to verify these 
  assumptions independently. Since the standard cosmological model allows for the presence of 
  small-scale inhomogeneities, and it only approaches the ideal CP asymptotically on large 
  scales, it is necessary to quantify the departure from the $\Lambda$CDM prediction as a function 
  of scale. In the late-time Universe this can be done by studying the fractality of the galaxy 
  distribution: in a pure fractal distribution, structures are found with the same amplitude on 
  arbitrarily large scales, and homogeneity is never reached. The reader is referred to 
  \citet{Martinez:2002}, and references therein, for a thorough introduction to the theory of 
  fractal point processes.
  
  Fractal dimensions quantify different moments of the counts-in-spheres in a point
  distribution. The most commonly used of them is the so-called correlation dimension $D_2(r)$,
  which quantifies the filling factor of spheres of different radii centred on points in the
  distribution. Using this kind of observables, different groups have been able to measure
  the transition from a fractal with $D_2<3$ to a homogeneous distribution $D_2=3$ on scales
  $r_H\sim100\,{\rm Mpc}$ \citep{1997NewA....2..517G, 2000MNRAS.318L..51P, 2001A&A...370..358K,
  2005ApJ...624...54H,2005BASI...33....1S,2009MNRAS.399L.128S,2012MNRAS.425..116S,
  2013MNRAS.434..398N}, while other authors claim that such transition has
  not yet been observed \citep{1999ApJ...514L...5J,2005A&A...443...11J,2011CQGra..28p4003S,
  2011EL.....9659001S,2014JCAP...07..035S}. In order to perform such analyses, full
  three-dimensional information for all the galaxies is necessary, and therefore
  these methods have only been used on spectroscopic catalogues, which traditionally cover
  much smaller volumes than their photometric counterparts. As we however showed in a
  previous publication \citep{2014MNRAS.440...10A}, it is possible to adapt this kind of
  study to 2-dimensional data projected on the celestial sphere and still be able extract
  information regarding the fractality of the local large-scale structure. This method can be
  combined with the traditional analysis of the scaling of the 1- and 2-point statistics
  of the galaxy distribution with magnitude limit to study the effective fractal dimension of
  the galaxy distribution using only angular information. With extra radial information,
  such as photometric redshifts, the constraints on possible departures from the CP can be
  enhanced further.
  
  In this work we apply these techniques to the 2MASS Photometric Redshift catalogue
  (2MPZ, \citealt{2014ApJS..210....9B}), an almost full-sky dataset providing comprehensive
  information on the galaxy distribution in the local Universe ($\bar{z}\sim0.1$). We are thus
  able to study possible departures from the CP on very large angular scales at late times,
  when these departures are expected to be maximal. The paper is structured as follows: in
  Section \ref{sec:theory} we present the methods and observables implemented on our data, as
  well as the $\Lambda$CDM predictions for these
  observables. Section \ref{sec:data} gives an overview of the 2MPZ catalogue, the sample used
  in this work and the criteria used in its selection. Section \ref{sec:results} presents our
  results regarding the fractality of the galaxy distribution in our sample. In Section
  \ref{ssec:systematics} we study the impact of different potential sources systematics that
  could affect our results. In particular we investigate the presence of hemispherical
  asymmetries in our final sample. Finally Section \ref{sec:discussion} summarizes
  our results. We also present, in Appendices \ref{app:mocks_lcdm} and \ref{app:mocks_fractal},
  the methods to generate mock $\Lambda$CDM and fractal
  realizations of the galaxy distribution that were used in this analysis.

\section{Tests of homogeneity using angular information}\label{sec:theory}
  Often in optical and near-infrared galaxy surveys it is not possible to measure
  precise redshifts for every object, mainly due to the large amount of time
  needed to integrate down the noise in a narrow-band spectrograph. Complete
  spectroscopic samples are thus typically shallower than photometric ones,
  reaching the shot noise limit faster. The situation is particularly striking in the context
  of \textit{all-sky} (4$\pi$ sterad) galaxy surveys. While the largest photometric catalogues
  covering the whole celestial sphere (optical SuperCOSMOS, infrared 2MASS and WISE) include
  hundreds of millions of sources, their largest spectroscopic counterpart, the 2MASS Redshift
  Survey \citep{2012ApJS..199...26H}, contains only 45,000 galaxies, hardly reaching beyond
  $z=0.05$. Deep all-sky catalogues are however essential if one desires to test the isotropy
  in matter distribution, and certainly favourable also for studying its homogeneity.
  
  If large volumes or number densities are needed for a particular study, it is often necessary
  to use datasets containing only angular coordinates and measured fluxes in a few wide bands
  for each object. Although this severely constrains the range of analyses that can be
  performed on these catalogues, there is still a great deal of cosmological
  information that can be extracted.
  In this Section we will describe different studies, related to the degree of homogeneity of
  the galaxy distribution, which can be performed using only angular information, starting with
  a brief review of angular clustering statistics.
  
  \subsection{Angular clustering}\label{ssec:th_clust}
    Probably the most informative observable regarding the statistics of the projected
    galaxy distribution is the angular two-point correlation function $w(\theta)$,
    defined as the excess probability of finding two galaxies with an angular
    separation $\theta$ with respect to an isotropic distribution
    \begin{equation}
      dP(\theta)=\bar{n}_{\Omega}^2\left[1+w(\theta)\right]\,d\Omega_1\,d\Omega_2,
    \end{equation}
    where $\bar{n}_{\Omega}$ is the mean angular number density of galaxies.

    The modelling of $w(\theta)$ has been extensively covered in the literature
    \citep{BookPeeblesLSS,2011MNRAS.414..329C}
    and we will only quote the main results here. The angular density of galaxies
    can be expanded in terms of its harmonic coefficients $a_{lm}$, which for a
    statistically isotropic distribution are uncorrelated and described by their
    angular power spectrum $C_l\equiv\langle|a_{lm}|^2\rangle$. The $C_l$ are
    straightforwardly related to $P_0(k)$, the 3D power spectrum at $z=0$:
    \begin{equation}\label{eq:cl2pk}
      C_l=\frac{2}{\pi}\int_0^\infty dk\,k^2\,P_0(k)|\omega_l(k)|^2,
    \end{equation}
    where
    \begin{equation}\label{eq:wincl}
      \omega_l(k)\equiv\int_0^\infty d\chi\,\chi^2\,W(\chi)\, G(z)\,
      [b(z)j_l(k\chi)-f(z)j_l''(k\chi)]
    \end{equation}
    Here $b(z)$ is the galaxy bias, $\chi$ is the radial comoving distance, related to the 
    redshift in a homogeneous background through
    \begin{equation}
      \chi(z)=\int_0^z\frac{c\,dz}{H(z)},
    \end{equation}
    $j_l(x)$ is the $l-$th order spherical Bessel function and $G(z)$ and $f(z)$
    are the linear growth factor and growth rate respectively (implicitly functions
    of the comoving distance $\chi$ on the lightcone). The quantity $W(\chi)$ above
    is the survey selection function, describing the average number density of sources
    as a function of the comoving distance to the observer and normalized to
    \begin{equation}
      \int_0^\infty W(\chi)\,\chi^2\,d\chi=1.
    \end{equation}
    Finally, the angular power spectrum is related to the angular correlation
    function through an expansion in Legendre polynomials $L_l$
    \begin{equation}
      w(\theta)=\sum_{l=0}^\infty\frac{2l+1}{4\pi}C_l\,L_l(\cos\theta).
    \end{equation}
    
    For small angular separations it is possible to use the so-called
    Limber approximation \citep{1953ApJ...117..134L}, which simplifies the relations
    above:
    \begin{align}\label{eq:limber2}
     &C_l=\int_0^\infty d\chi
     \left[\chi\,W(\chi)\,b(z)\,G(z)\right]^2P\left(k=\frac{l+1/2}{\chi}\right),\\
     \label{eq:limber1}
     &w(\theta)=\int_0^2d\chi\,\chi^4\,W^2(\chi)\int_{-\infty}^\infty d\pi\,
     \xi\left(\sqrt{\pi^2+\chi^2\theta^2}\right),
    \end{align}
    where $\xi(r)$ is the three-dimensional correlation function, and $\pi$ is the
    radial separation between two galaxies.

  \subsection{The angular homogeneity index}\label{ssec:th_htheta}
    In a three-dimensional framework, one of the most commonly used observables to describe
    the fractality of a point distribution is the so-called correlation dimension. Let us
    first define the correlation integral $C_2(r)$ as the average number of points contained
    by spheres of radius $r$ centred on other points of the distribution. For an infinite
    homogeneous point process this quantity would grow like the volume $C_2(r)\propto r^3$.
    The correlation dimension is thus defined as the logarithmic tilt of the correlation
    integral:
    \begin{equation}
      D_2(r)\equiv\frac{d\log C_2}{d\log r},
    \end{equation}
    and hence, if the point distribution is uncorrelated on large scales, $D_2(r)$ should
    approach $3$ for large $r$. In a FRW universe we can expect deviations from this value
    due to the gravitational clustering of density perturbations\footnote{The finiteness of
    the point distribution will also cause deviations from homogeneity due to shot noise.
    This can be fully incorporated in the modelling of $D_2$ and $H_2(\theta)$. See
    \citet{2014MNRAS.440...10A} for further details.}, but the homogeneous result should be
    approached asymptotically on large scales according to the Cosmological Principle.
    See \citet{Bagla:2007tv,2010MNRAS.405.2009Y} for a thorough modelling of these quantities
    within the standard cosmological model.
    
    In our case, due to the lack of precise radial information, we will use instead the angular
    homogeneity index $H_2(\theta)$, defined in \citet{2014MNRAS.440...10A} by directly
    adapting the definitions of $C_2$ and $D_2$ to a 2-dimensional spherical space. 
    That is, we define the angular correlation integral $G_2(\theta)$ as the average
    counts of objects in spherical caps (instead of spheres), and the angular
    homogeneity index $H_2(\theta)$ as the tilt of $G_2$ with the area of these spherical
    caps $V(\theta)\equiv 2\pi(1-\cos\theta)$. Thus $H_2$ is normalized to be $1$ for
    an infinite homogeneous distribution, and in a FRW universe it can be related to
    the angular correlation function (to first order) via
    \begin{equation}
     H_2(\theta)=1-\frac{\bar{w}(\theta)-w(\theta)}{1+\bar{w}(\theta)}-
     \frac{1}{2\pi\bar{n}_\Omega\,(1-\cos\theta)},
    \end{equation}
    where
    \begin{equation}
     \bar{w}(\theta)\equiv\frac{1}{1-\cos\theta}\int_0^\theta w(\theta)\sin\theta\,d\theta.
    \end{equation}
    
    We elaborate further on the technical details regarding the estimation of the angular
    homogeneity index from a galaxy survey and how to deal with boundary effects
    in Section \ref{ssec:htheta}. These are also more thoroughly discussed in
    \citet{2014MNRAS.440...10A}, and we refer the reader to that paper for further
    information.

    One may think that the net effect of using only angular information directly translates
    in a decrease of sensitivity, but this is not necessarily the case, especially when the
    aim is to provide model-independent constraints. Using only angular, and therefore
    observable information has the advantage that no assumption about the underlying
    cosmological model is necessary in order to convert redshifts into distances. This
    therefore allows for a cleaner test of homogeneity. Other probes, although more
    sensitive a priori, are ultimately a consistency test of some classes of cosmological
    models.

  \subsection{Scaling relations}\label{ssec:th_scaling}
    As pointed out by \citet{BookPeeblesCosmo}, a lot of information regarding the degree
    of homogeneity of the galaxy distribution can be extracted from the scaling of
    different observables with the limiting flux for a magnitude-limited survey.
    Given the sample's luminosity function $\phi(L,z)$, describing the number density
    of galaxies in an interval $dL$ of luminosity at redshift $z$, a survey with a
    limiting flux $F_c$ should observe a number density of galaxies as a function of
    distance $\chi$ given by
    \begin{equation}
      \bar{n}(\chi,>F_c)=\int_{L_{\rm min}}^{\infty}\phi(L,z(\chi))dL,
    \end{equation}
    where $L_{\rm min}=4\pi\,F_c\,d_L^2(\chi)$ is the limiting luminosity at a distance
    $\chi$ and $d_L$ is the luminosity distance. For low redshifts we can approximate
    $d_L\sim \chi$, and hence, for a non-evolving population ($\phi(L,z)\equiv\phi(L)$)
    we obtain that the number density should be purely a function of the combination
    $\sqrt{F_c}\chi$:
    \begin{equation}
      \bar{n}(\chi,>F_c)\equiv g(x\equiv\sqrt{F_c}\chi)=\int_{4\pi x^2}^\infty\phi(L)\,dL.
    \end{equation}
    
    Without any redshift information we will actually observe the projected number 
    density of galaxies, defined as the number of galaxies observed per unit solid angle:
    \begin{equation}
      \bar{n}_{\Omega}(>F_c)=\int_0^\infty \bar{n}(\chi,>F_c)\,\chi^2\,d\chi,
    \end{equation}
    from which it is easy to extract the scaling law:
    \begin{equation}\label{eq:scaling_dens}
      \bar{n}_{\Omega}(>F_c)\propto F_c^{-3/2}\propto 10^{\beta\,m_{\rm lim}},
    \end{equation}
    where $m_{\rm lim}$ is the apparent magnitude limit and $\beta=0.6$. 
    Although it is not easy to assess the expected observational uncertainty on $\beta$, 
    as shown in~\citet{BookPeeblesCosmo}, fractal models would predict a much smaller value. 
    In particular it is easy to prove that for a model with a fractal dimension $D$, in which
    number counts follow the law $N\propto R^D$, we would measure $\beta\sim0.2\,D$.
    
    An even stronger relation can be found for the two-point angular correlation
    function. Using Eq. \ref{eq:limber1} and the fact that
    $W(\chi)\propto\bar{n}(\chi,>F_c)$, it is straightforward to show that for two
    different flux cuts $F_1$ and $F_2$ the corresponding correlation functions would be
    related by
    \begin{equation}\label{eq:scaling_wth}
     w_2(\theta)=\frac{w_1(B\theta)}{B},
    \end{equation}
    where the scaling factor $B$ is\footnote{This scaling factor has
    traditionally been labelled $D$ instead of $B$, but we use a different convention here
    to avoid any confusion with the fractal dimension.} $B\sim\sqrt{F_1/F_2}$.
    
    This is a well known result: as we increase the depth of the survey, we increase the
    chance that pairs of galaxies that are distant from each other (and therefore
    uncorrelated) will subtend small angles, thus decreasing the amplitude of the angular
    correlation function. Traditionally the scaling relation above has been used as a
    tool to rule out systematic errors associated with incorrect angular masking, however
    it can also be used as a consistency check to verify the statistical homogeneity of the
    galaxy distribution.

    The reason for this can be understood intuitively. Consider a perfect fractal
    distribution, for which structures are found on all scales with the same amplitude. As
    we increase the survey depth, we will also include larger and larger structures, an
    effect which compensates for the loss of correlation described above. The result is
    that for a scale-independent fractal, the angular correlation function is independent
    of the survey depth, and hence of the magnitude limit (see \citealt{BookPeeblesCosmo}
    for a precise derivation of this result in the case of Rayleigh-Levy flights). This
    test complements the calculation of the angular homogeneity index, described in the
    next Section, in that it is able to probe the degree of homogeneity also in the radial
    direction.
    
    Although the measurement of these scaling relations is an important test of homogeneity
    and isotropy, these analysis suffer from a number of caveats that must be highlighted.
    First of all, assuming a non-evolving luminosity function is not necessarily correct
    for all galaxy populations, and could induce a bias in our measurement of $\beta$.
    Secondly, while nothing prevents us from computing the two-point correlation function
    in an inhomogeneous dataset, its interpretation in terms of an excess probability of
    finding pairs of galaxies is only valid in the homogeneous case. Because of this,
    the study of the scaling relations should be undestood as a consistency test of the
    homogeneous model, and not as a model-independent constraint on inhomogeneous
    cosmologies. Finally, it is worth noting that the exact form of the scaling
    relations presented here should vary in deeper catalogues reaching larger
    redshifts, which will be a concern for future surveys. This is, however, not a
    issue for 2MPZ, where $\bar{z}\sim0.1$.

\section{The data}\label{sec:data}
  \subsection{The 2MASS Photometric Redshift catalogue}\label{ssec:2MPZ}
    
    The 2MASS Photometric Redshift catalogue (2MPZ, \citealt{2014ApJS..210....9B}) is the first
    publicly available\footnote{Available for download from
    \url{http://surveys.roe.ac.uk/ssa/TWOMPZ}.}  all-sky dataset that provides photometric
    redshift information. Its parent sample, the 2MASS Extended Source Catalogue (XSC,
    \citealt{2000AJ....119.2498J}) includes over 1.6 million resolved sources (mostly galaxies),
    detected on most of the sky except for the highly confused Galactic Bulge, and provides
    precise astro- and photometric information, the latter in three near-infrared (IR) bands,
    $J$, $H$ and $K_s$. About 1 million of the 2MASS galaxies are within its approximate
    completeness limit of $K_s \lesssim 13.9$ mag (Vega). By cross-matching the 2MASS XSC
    sample with two other major all-sky photometric surveys (deeper than 2MASS), SuperCOSMOS
    scans of photographic plates \citep{2001MNRAS.326.1279H} and mid-IR satellite data from
    WISE \citep{2010AJ....140.1868W}, \citet{2014ApJS..210....9B} obtained multiwavelength
    information for the majority (95\%) of 2MASS galaxies. This allowed further to derive
    photometric redshifts for these sources, by employing the empirical ANNz algorithm
    \citep{2004PASP..116..345C} trained on subsamples drawn from spectroscopic redshift surveys
    overlapping with 2MASS. The 2MPZ is an extension of earlier attempts by
    \citet{2004PASA...21..396J} and \citet{2010MNRAS.406....2F} who derived less accurate
    photo-$z$'s for 2MASS, not having access to the WISE data collected in 2010. The median
    redshift of the 2MPZ sample is $\bar{z}=0.08$ and its typical photo-$z$ errors are 13\%
    (RMS in $\delta z$ of $\sim 0.013$).
    
  \subsection{Sample selection}\label{ssec:sample}    
    Even though 2MASS has virtually 100\% sky coverage, a number of observational effects
    will inevitably reduce the fraction of the sky that can be used for cosmological
    studies. The most important of these result from our Galaxy obscuring the view and
    creating the so-called Zone of Avoidance. In addition, as 2MPZ was built by
    cross-correlating the 2MASS XSC  with WISE and SuperCOSMOS, the two latter catalogues
    bring their own additional incompletenesses, which need to be accounted for. In order
    to determine the regions of the sky that should be masked for the subsequent analysis,
    we have performed a standard study of source number counts in the presence of the
    different possible systematics.
    For the 2MASS XSC, and hence 2MPZ, there exist 4 potential sources of systematic
    effects: Galactic dust extinction, stars, seeing and sky brightness. The effects of
    each of these  on the angular correlation function of 2MASS galaxies were
    studied by \citet{2005ApJ...619..147M}, who found that both sky brightness and seeing
    had a negligible effect. We have therefore focused our analysis on dust extinction
    and star density.
    
    For a particular region in the sky, the contamination due to Galactic dust can be
    quantified in terms of the $K$-band correction for Galactic extinction,
    $A_K=0.367\,E(B-V)$, where the reddening $E(B-V)$ was derived for the whole
    sky by \citet{1998ApJ...500..525S}. Star density in turn was computed
    at the position of each 2MASS extended sources based on local counts of point sources
    (stars) brighter than $K_s=14$ mag \citep{2006AJ....131.1163S}. This information (log star
    density per deg$^2$) is provided for every source in the public 2MPZ database, and can be
    used to create a full-sky map of $n_{\rm star}$.  We generated maps of $A_K$ and
    $n_{\rm star}$ using the HEALPix\footnote{\url{http://healpix.sourceforge.net/}}
    pixelization scheme \citep{2005ApJ...622..759G} with a resolution parameter
    ${\tt Nside}=64$ (pixels of $\delta\Omega\simeq0.84\,{\rm deg}^2$). This is the fiducial
    resolution that was used for most of this work, unless otherwise stated.
  
  \begin{figure*}
    \centering
    \includegraphics[width=0.48\textwidth]{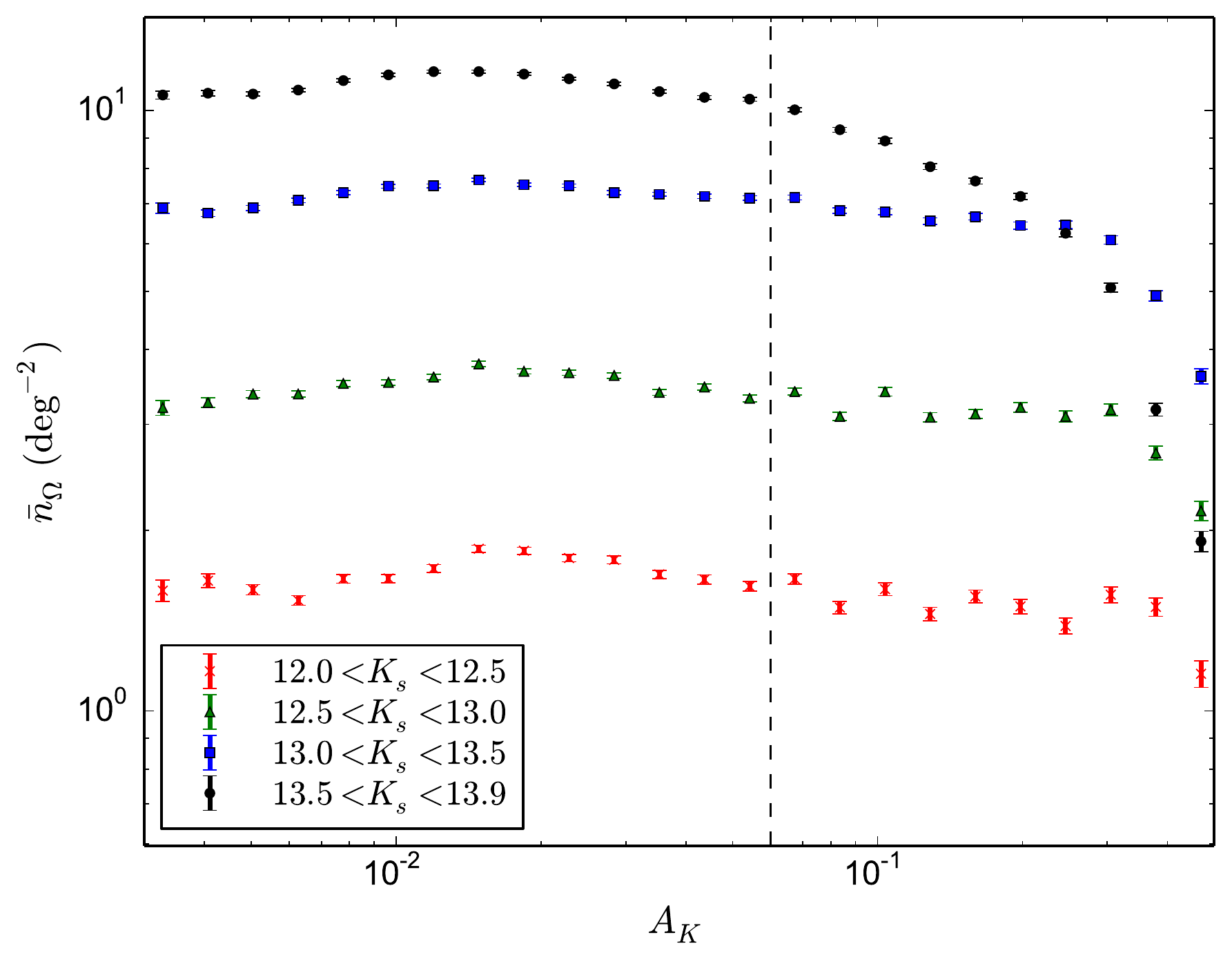}
    \includegraphics[width=0.49\textwidth]{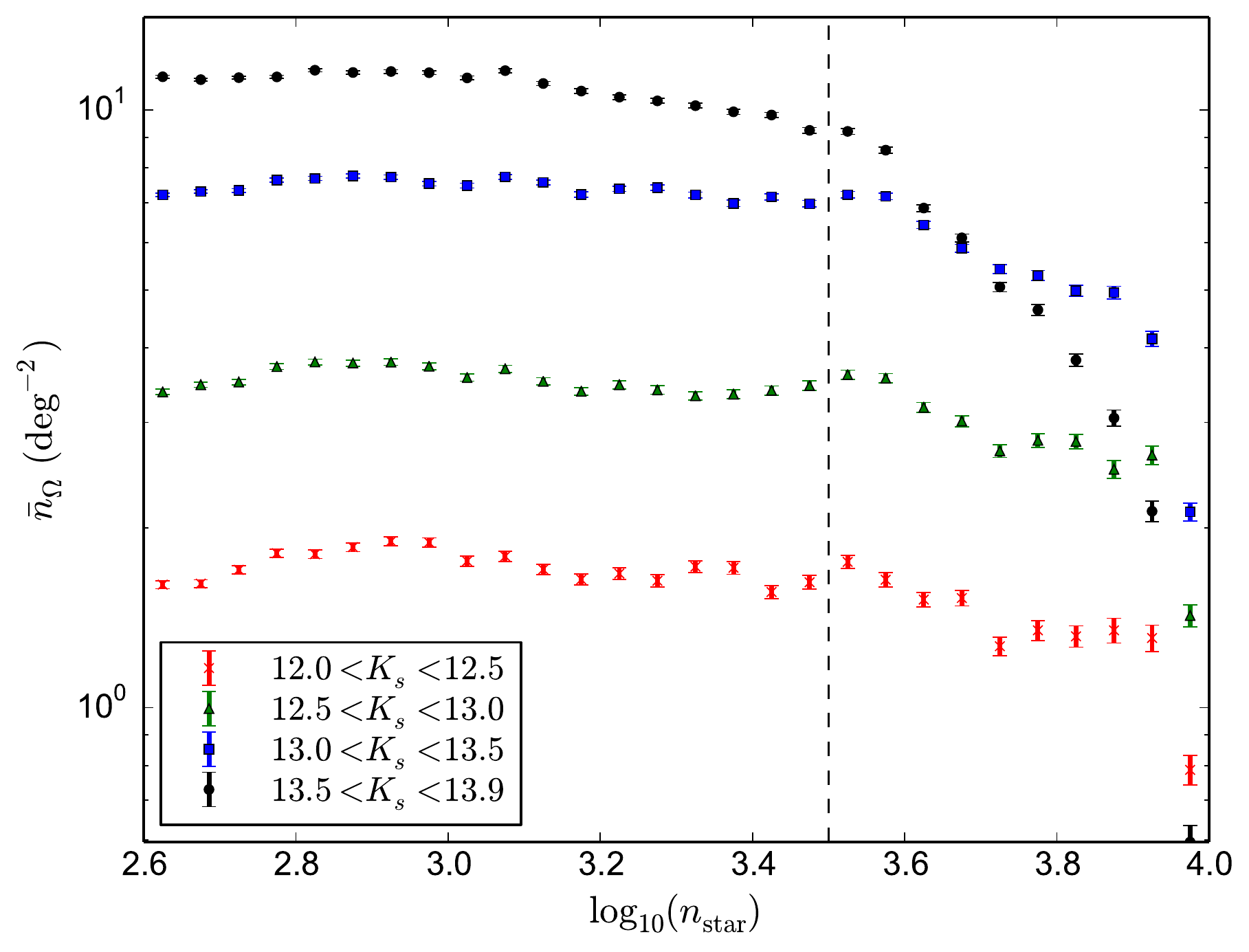}
    \caption{Number density of 2MPZ galaxies in 4 bins of magnitude measured in pixels with
             varying dust extinction (left panel) and star density (right panel). The vertical
             dashed lines show the values of $A_{K,{\rm max}}$ and $n_{\rm star,max}$ chosen
             to avoid these two systematics.}
    \label{fig:mask_ndens}
  \end{figure*}  
  
  \begin{figure*}
    \centering
    \includegraphics[height=0.90\textheight]{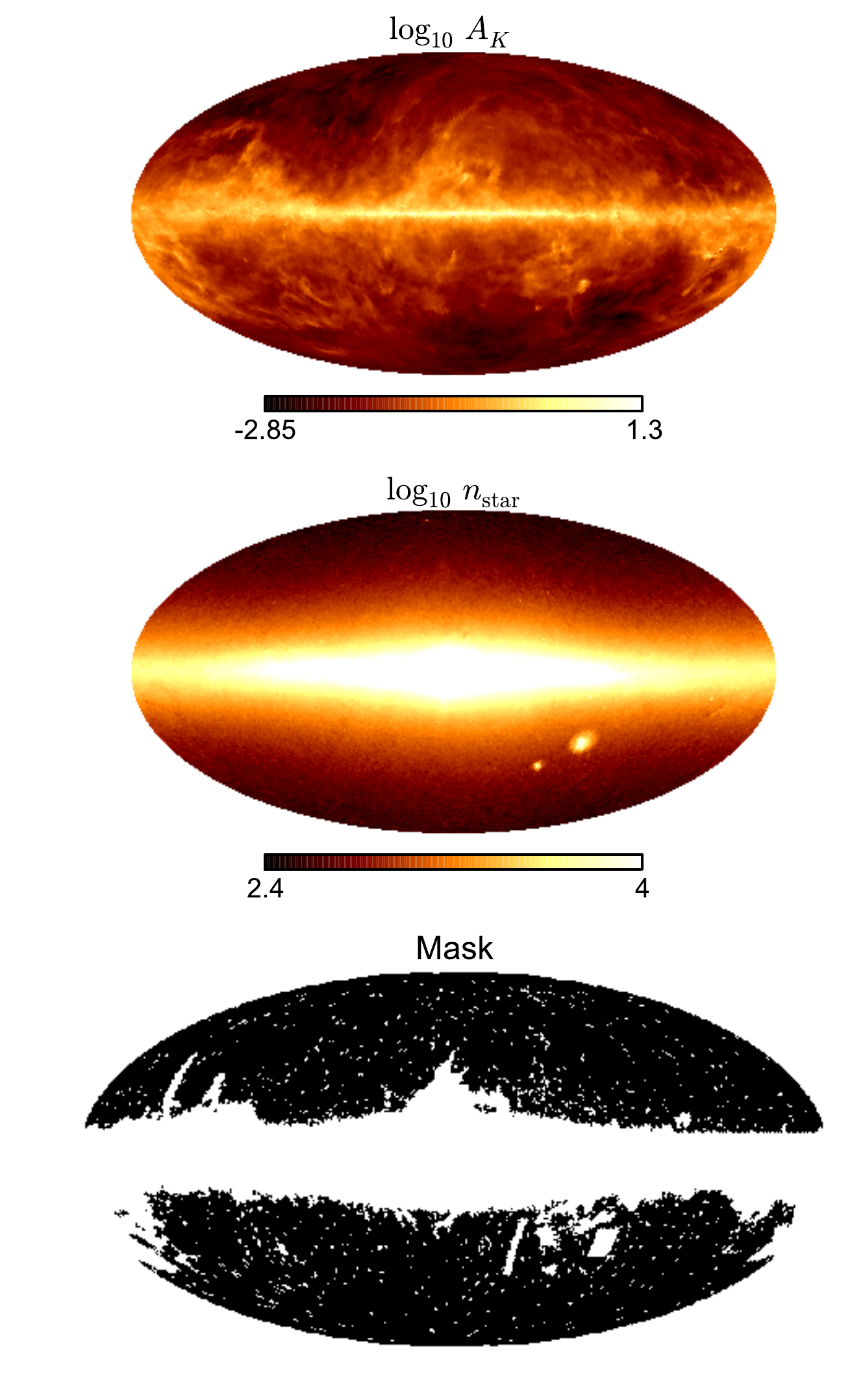}
    \caption{Sky maps of the two main sources of systematics: dust extinction (upper panel)
             and star density (middle panel). Our final mask is shown in the lower panel, and
             was defined by minimizing the effects of these systematics as described in 
             Sections \ref{ssec:sample} and \ref{ssec:systematics}.}
    \label{fig:maps1}
  \end{figure*}
   
    Following an analysis similar to \citet{2004PhRvD..69h3524A} we computed the
    angular number density of sources for objects with different magnitude
    $K_s$ and residing in pixels with a different value of $A_K$ and $n_{\rm star}$, and
    selected threshold values $A_{K,{\rm max}}$ and $n_{\rm star,max}$ as those beyond
    which a substantial decrease in the observed number densities was observed. The results
    are shown in Figure \ref{fig:mask_ndens} for dust extinction (upper panel) and star
    density (lower panel). In view of this result we chose the thresholds $A_{K,{\rm max}}
    =0.06$ and $\log_{10}(n_{\rm star,max})=3.5$, which eliminates the areas near the
    Galactic plane and at the Magellanic Clouds (the latter were additionally cut out manually
    for better completeness), reducing the usable sky fraction to about $69\%$.
    Besides this, a small subset of pixels had to be discarded due to incompleteness in WISE
    and SuperCOSMOS. The former, in its 'All-Sky' release used for 2MPZ construction, is
    incomplete in two strips at Ecliptic $\lambda,\beta = 100^\circ,+45^\circ$ and
    $290^\circ,-45^\circ$ due to so-called "torque rod gashes"\footnote{
    \url{http://wise2.ipac.caltech.edu/docs/release/allsky/expsup/sec6\_2.html\#lowcoverage}};
    these were masked out manually. In the latter, a small fraction of data were missing due
    to issues with "stepwedges", which affected mostly plate corners; these create a regular
    pattern near the equator and were identified by comparison of the parent 2MASS XSC
    dataset with the final 2MPZ sample. The same comparison allowed us also to identify
    other sources of incompleteness brought about by WISE and SuperCOSMOS, which is mostly
    saturation around the brightest stars. The final footprint used for the best part of our
    analysis is shown in Figure \ref{fig:maps1} together with the maps of $A_K$ and
    $n_{\rm star}$, and covers $f_{\rm sky}=0.647$.
    
    For most of this work we used a fiducial sample of galaxies with $12.0<K_s<13.9$. The
    lower magnitude cut was chosen to slightly reduce the number of local structures that
    could complicate the analysis, as well as the interpretation of the results. The upper
    cut was in turn estimated by \citet{2014ApJS..210....9B} to give a uniform sky coverage.
    It is worth noting that we verified that slight variations in these magnitude limits
    did not vary our results significantly. Taking into account the mask described above, our
    fiducial sample contains 628280 galaxies. In the analysis of the homogeneity index
    (Section \ref{ssec:htheta}), we have further divided the survey into two photometric
    redshift bins, with $0.03\leq z_{\rm ph}<0.08$ and $0.08\leq z_{\rm ph}\leq0.3$, each
    containing 264158 and 351383 objects respectively. From here on we will refer to these
    subsamples as ``Bin 1'' and ``Bin 2''. The number density field corresponding to our
    fiducial sample is displayed in the upper panel of Figure \ref{fig:maps2}.
    
    In order to verify that the cuts defining our sample do not introduce any systematic
    biases in our results we have studied their effect on the 2-point clustering statistics
    as well as the presence of hemispherical asymmetries in our final density maps. This
    is described in Section \ref{ssec:systematics}.

  \subsection{Clustering analysis and galaxy bias}\label{ssec:bias}
    In order to characterize the possible deviations from statistical homogeneity that we will
    study in the next Section, it is necessary to compare them with the expected statistical
    uncertainties allowed within the standard cosmological model. The most reliable way of
    estimating these is by using mock galaxy catalogues that reproduce the statistical properties
    of our survey. For this we have used the method described in Appendix \ref{app:mocks_lcdm},
    which requires a correct model of the best-fit angular power spectrum of the data.

    \begin{table*}
      \begin{center}
        \begin{tabular}{l|c|c|c|c|c|c}
          Sample name & Cuts & $N_{\rm gal}$ & $\alpha$ & $\beta$ & $z_0$ & Bias $b$\\
          \hline
          Fiducial & $K_s\in[12,13.9]$                               & 628280 & 
          2.21 & 1.43 & 0.053 & $1.24\pm0.03$ \\
          Bin 1    & $K_s\in[12,13.9]$ \& $z_{\rm ph}\in[0.03,0.08)$ & 264158 & 
          2.61 & 3.36 & 0.066 & $1.18\pm0.03$ \\
          Bin 2    & $K_s\in[12,13.9]$ \& $z_{\rm ph}\in[0.08,0.3]$  & 351383 & 
          6.51 & 1.30 & 0.032 & $1.52\pm0.03$ \\
          \hline
        \end{tabular}
      \end{center}
      \caption{Summary of the galaxy samples used in this work. Column 2 lists the cuts imposed
               in each case (besides those used to define the mask), and column 3 lists the
               total number of objects in each sample. Columns 4-6 contain the best-fit
               parameters for the redshift distributions according to the model in Eq.
               \ref{eq:nz_model}, and column 7 shows the value of the best-fit galaxy bias.}
      \label{tab:nz_bf}
    \end{table*}
    \begin{figure}
      \centering
      \includegraphics[width=0.49\textwidth]{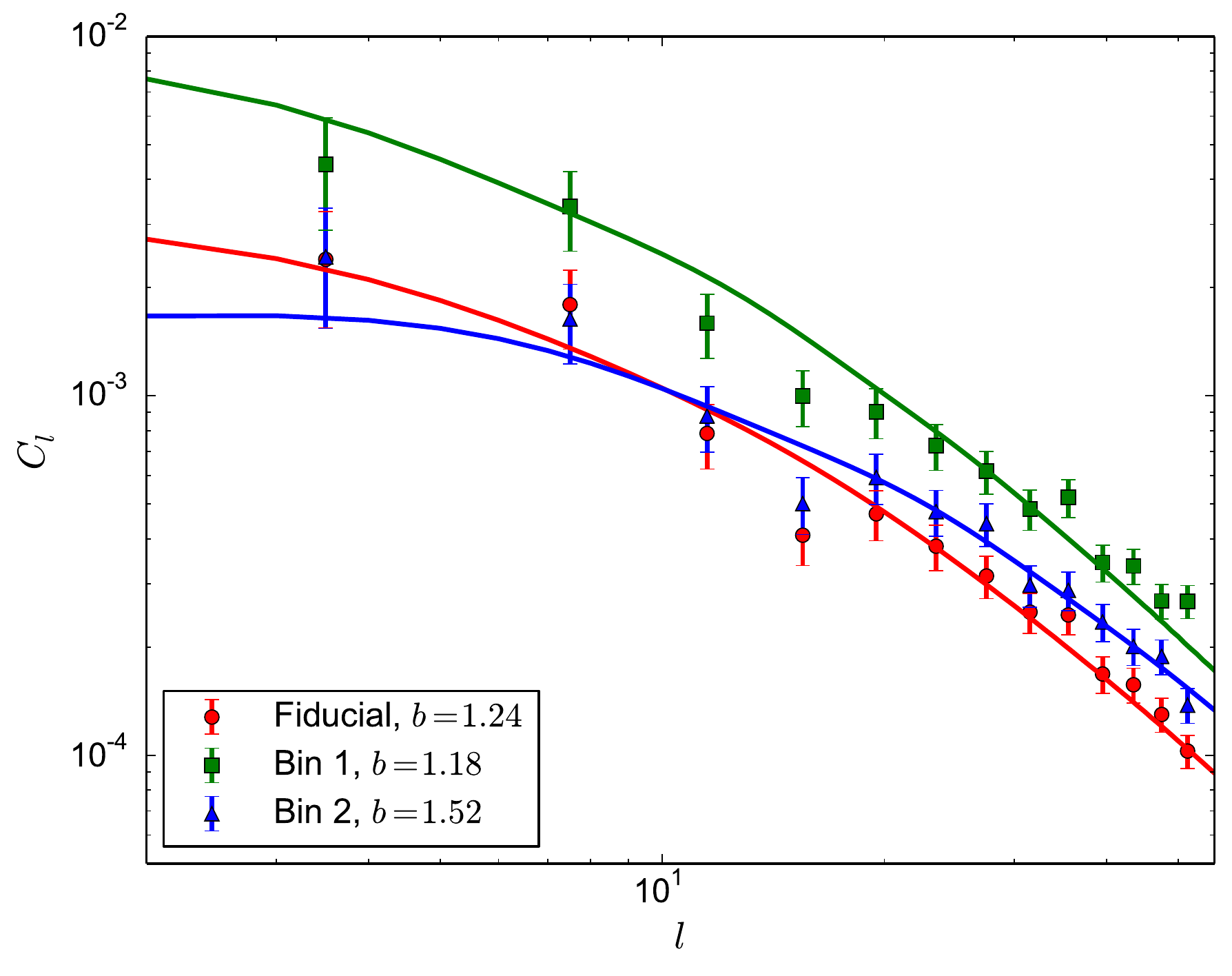}
      \caption{Angular power spectrum measured for the three samples listed in table
               \ref{tab:nz_bf} (points with error bars) together with the $\Lambda$CDM
               prediction (solid lines) using the best-fit bias parameters.}
      \label{fig:bias_fit}
    \end{figure}
    \begin{figure}
      \centering
      \includegraphics[width=0.49\textwidth]{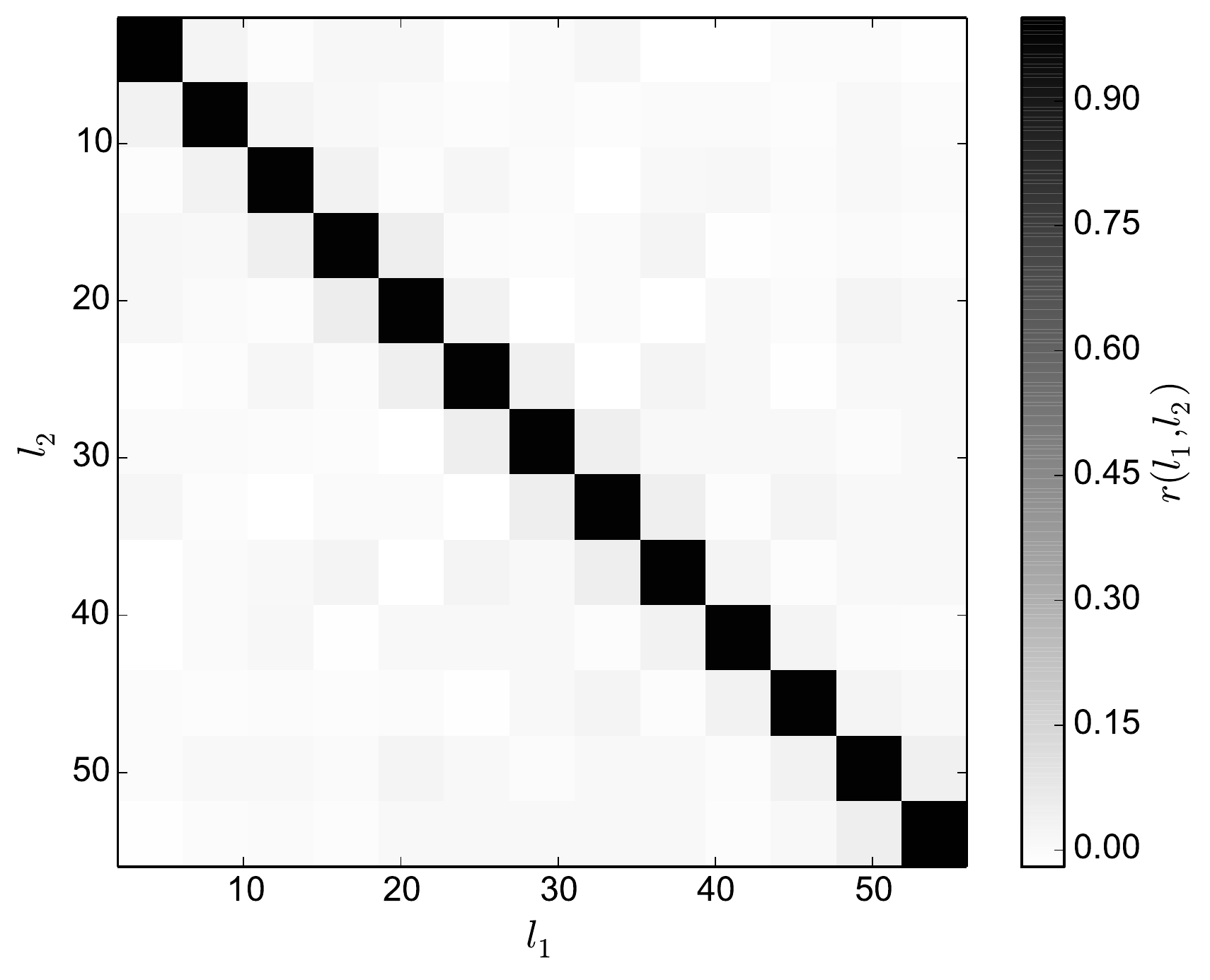}
      \caption{Correlation matrix ($r_{ij}\equiv\mathcal{C}_{ij}/\sqrt{\mathcal{C}_{ii}\mathcal{C}_{jj}}$)
               of the uncertainties in the angular power spectrum for our fiducial sample. The measurements
               in different bins of $l$ are almost completely uncorrelated.}
      \label{fig:corrmat}
    \end{figure}

    A crucial step in characterizing the clustering statistics of our galaxy sample within
    $\Lambda$CDM is modelling its redshift distribution $dN/dz$. Although 2MPZ provides
    photometric redshifts for all the sources with a remarkably small uncertainty (on average),
    it is not possible to estimate the true $dN/dz$ reliably using these: $dN/dz_\mathrm{phot}$
    is a convolution of the underlying $dN/dz_\mathrm{spec}$ with the photometric redshift
    error, which typically makes the photo-$z$ distribution narrower than the true one.   
    Fortunately, at high Galactic latitudes $b\gtrsim60^\circ$ there is practically full
    spectroscopic coverage from SDSS\footnote{2MPZ used SDSS Data Release 9
    \citep{SDSS.DR9}.}, which we can use for this task\footnote{Spectroscopic redshifts, where
    available, are also provided in the 2MPZ database.}. In total we identified a subset of
    $\sim 10^5$ objects in this region with spectroscopic redshifts measured by SDSS, which we
    binned to estimate the redshift distributions of our galaxy samples. We fit a smooth function
    of the form
    \begin{equation}\label{eq:nz_model}
      \frac{dN_{\rm model}}{dz}\propto z^\alpha\,\exp\left[-\left(\frac{z}{z_0}\right)^\beta\right]
    \end{equation}
    to each of these histograms, obtaining the best-fit parameters listed in Table \ref{tab:nz_bf}.
    In order to verify that the redshift distribution estimated from this spectroscopic sample
    can be extrapolated to the rest of the survey, we calculated the angular two-point correlation
    function in this region and in the whole survey for our fiducial sample and compared both of
    them. A difference in the redshift distributions would cause a difference in the amplitude of
    the correlation functions, which we did not observe.
    
    For each of the three subsamples listed above (Fiducial, Bin 1 and Bin 2) we determined a
    single effective bias parameter $b$ that best fits its clustering statistics. To do this
    we first created a map of the projected overdensity of galaxies with the angular resolution
    parameter ${\tt Nside}=64$ by assigning to each pixel $i$ the value $\delta_i=N_i/\bar{N}-1$,
    where $N_i$ is the number of galaxies in that pixel and $\bar{N}$ is the average number
    of galaxies per pixel. We then computed the angular power spectrum $C_l$ of this overdensity
    field, fully accounting for the angular mask using the PolSpice software package
    \citep{2004MNRAS.350..914C}. The theoretical power spectrum we fitted to these data was
    calculated using the fits to the redshift distribution described above as proxy for the
    radial window function ($W(\chi)\chi^2d\chi/dz\propto dN/dz$) and Equations \ref{eq:cl2pk}
    and \ref{eq:wincl}, with the linear power spectrum at $z=0$, $P_0(k)$, predicted by CAMB
    \citep{2000ApJ...538..473L}. We fixed all cosmological parameters except for the linear galaxy
    bias to their best-fit values as measured by \citet{2014A&A...571A..16P}, $(\Omega_M,
    \Omega_\Lambda,\Omega_b,h,\sigma_8,n_s)=(0.315,0.685,0.049,0.67,0.834,0.96)$. The best-fit
    value of the bias $b$ was found by minimizing the $\chi^2$:
    \begin{equation}
      \chi^2\equiv\sum_{l_1,l_2}[\hat{C}_{l_1}-C_{l_1}^{\rm model}(b)]\,\mathcal{C}^{-1}_{l_1l_2}
                                [\hat{C}_{l_2}-C_{l_2}^{\rm model}(b)],
    \end{equation}
    where $\hat{C}_l$ is the measured power spectrum, $C_l(b)$ is the linear $\Lambda$CDM model
    for a bias $b$ and $\mathcal{C}_{l_1l_2}$ is the covariance between different multipoles.
    
    For this exercise we assumed a diagonal covariance matrix, estimated theoretically as
    \citep{2010MNRAS.406....2F,2011MNRAS.414..329C}:
    \begin{equation}
      \mathcal{C}_{l_1l_2}=\frac{2}{f_{\rm sky}(2l_1+1)}
      \left(\hat{C}_{l_1}+\frac{1}{\bar{n}_\Omega}\right)\delta_{l_1l_2}.
    \end{equation}
    This assumption is exact for a statistically isotropic distribution observed across the whole
    sky, however it is well known that partial sky coverage introduces correlations between
    different multipoles. In order to avoid this we grouped the multipoles into bins of width
    $\Delta l=4$, thus reducing the correlation between neighbouring bins. A posteriori we
    also confirmed the validity of this approximation by computing the full covariance 
    matrix $\mathcal{C}_{l_1l_2}$ from 10000 mock catalogues generated using our best-fit
    values for the bias $b$ (see Figure \ref{fig:corrmat}) . It is also worth noting that in order
    to avoid using an incorrect model for the angular power spectrum or its covariance matrix
    due to small-scale non-linearities we limited the range of multipoles used for this analysis
    to $l\in[2,56]$, corresponding to scales $k\lesssim0.3\,h{\rm Mpc}^{-1}$.
    
    The best-fit values of $b$ in the three samples are listed in the last column of
    Table \ref{tab:nz_bf}, and are in good agreement with prior estimates of the bias for
    2MASS galaxies \citep{2003ApJ...598L...1M,2005MNRAS.364..593F,2010MNRAS.406....2F}. 
    Figure \ref{fig:bias_fit} shows the power spectra computed in our three samples together
    with their best-fit theoretical predictions. It is worth noting that these estimates
    rely heavily on the assumed amplitude of the dark matter power spectrum (i.e. $\sigma_8$),
    and therefore our results must be understood as measurements of the combination
    $b\times(\sigma_{8,{\rm Planck}}/\sigma_8)$, with $\sigma_{8,{\rm Planck}}=0.834$.

\section{Results}\label{sec:results}
  This Section presents in detail our analysis regarding the fractality of the galaxy
  distribution in our sample. Sections \ref{ssec:scaling} and \ref{ssec:htheta} discuss
  respectively the scaling relations and the angular homogeneity index derived for the 2MPZ
  sample, while in Section \ref{ssec:systematics} we analyse the possible
  systematic effects that could have an impact on these results, including the presence
  of significant hemispherical or dipolar asymmetries.

  \subsection{Scaling relations}\label{ssec:scaling}
    As a means to probe the possible fractal structure of the galaxy distribution we have
    studied the scaling of the number density and angular correlation function of
    galaxies with magnitude limit and verified the predictions described in Section
    \ref{ssec:th_scaling} for homogeneous cosmologies.
    
    In order to form a quantitative idea regarding the agreement of our results with the
    standard cosmological model we have compared the results obtained from the 2MPZ
    catalogue with those of a suite of mock realizations of the fractal $\beta$-model,
    described in Appendix \ref{app:mocks_fractal}. These mocks were made to follow the
    $K_s$-band luminosity function estimated by \citet{2014JCAP...10..070A}, and span a
    physical volume equivalent to that probed by 2MPZ. Note that the luminosity function
    used to simulate the fractal realizations was derived from the 2MPZ data assuming a
    FRW background, and it is therefore inconsistent in principle to apply it to the
    inhomogeneous fractal models. However, since we only aim to use the scaling relations
    as a consistency test of the CP, the use of this luminosity function does not affect
    our final results. We generated 100 mock realizations for 3 different values of the
    fractal dimension: $D=2.5,\,2.75$ and 2.90.
    
    \subsubsection{Number density}\label{sssec:scaling_ndens}
      \begin{figure}
        \centering
        \includegraphics[width=0.49\textwidth]{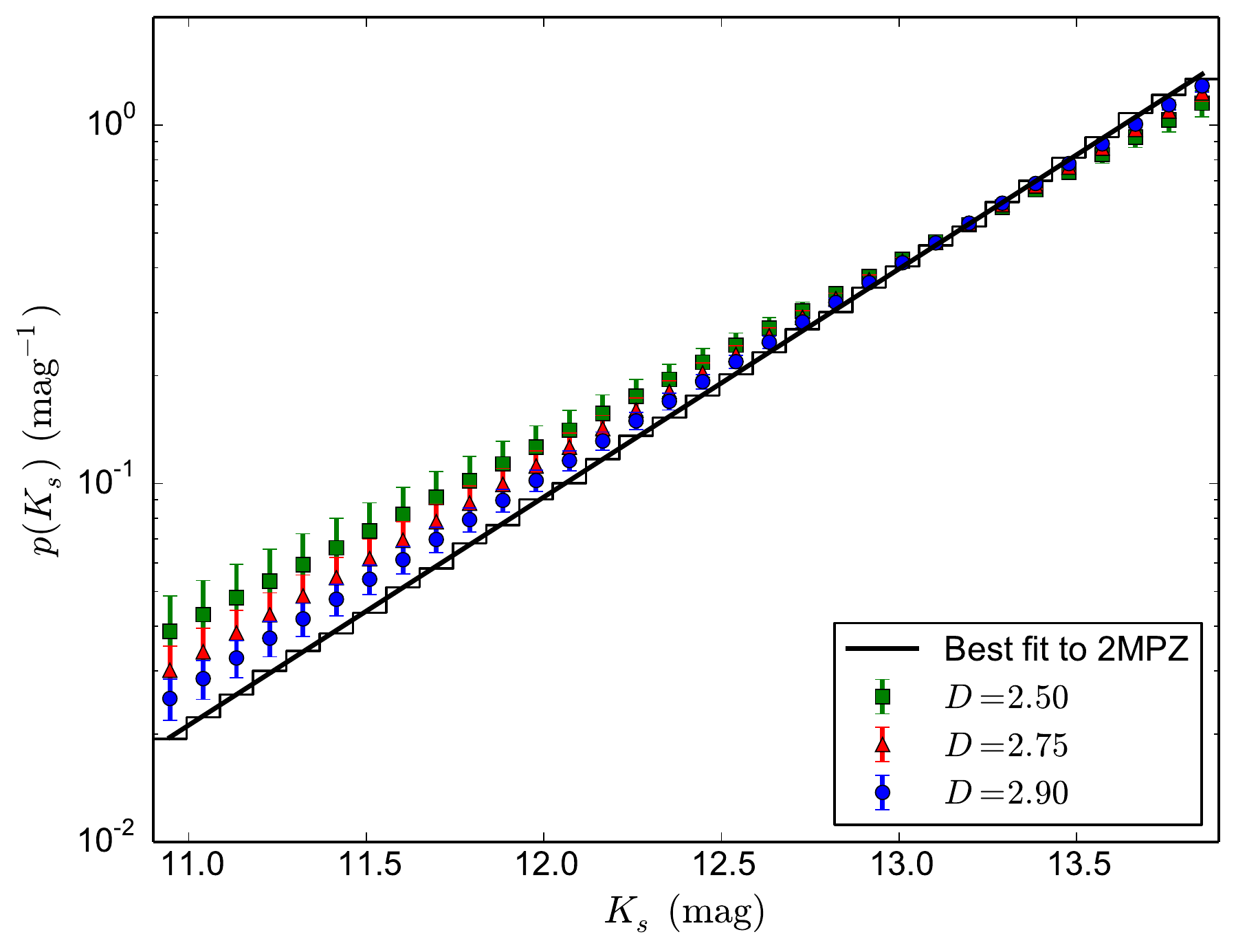}
        \caption{Probability distribution for $K_s$ derived from the number counts
                 histograms for fractal models with different fractal dimensions
                 (points with error bars) together with the best-fit 
                 parametrization $\beta=0.63$ for the 2MPZ sample (black solid line).
                 The black histogram shows the actual 2MPZ data.}
        \label{fig:scaling_ndens}
      \end{figure}
      As shown in Section \ref{ssec:th_scaling}, for a homogeneous Universe the number
      of galaxies observed by a survey with magnitude limit $m_{\rm lim}$ should follow
      an exponential distribution given by equation \ref{eq:scaling_dens}. Consequently
      the probability distribution function of magnitudes should follow a similar
      behaviour with the same index $\beta=0.6$:
      \begin{equation}\label{eq:linmod}
        \log_{10} {\rm pdf}(m)=\beta\,m+C.
      \end{equation}
      Deviations with respect to $\beta=0.6$
      can be expected within the standard model due to clustering variance and shot noise,
      and the exact uncertainty on this parameter is difficult to assess without an accurate
      model of the three-dimensional distribution of galaxies.
      
      In order to estimate $\beta$ from the data we made a histogram of the number counts
      of galaxies as a function of apparent magnitude in the $K_s$ band. The value of $\beta$
      was computed by fitting the logarithmic number counts to the linear model in Eq.
      \eqref{eq:linmod}. Doing this we obtained a best-fit value $\beta=0.63$, which is
      $5\%$ above the homogeneous value. Understanding the significance of this deviation
      requires computing the uncertainty on this parameter, which is related to the variance
      of the number counts. As we discuss further in the context of hemispherical asymmetries
      (Section \ref{sssec:sys_hemisph}), the errors in the number counts are dominated by cosmic
      variance, and are a factor of $\sim12.5$ larger than the Poisson errors ($\propto\sqrt{N}$)
      for our fiducial sample. In order to estimate the error of $\beta$ we have therefore used
      the Poisson error of the histogram scaled by a constant factor of $12.5$,
      obtaining $\Delta\beta\simeq0.015$, i.e. our measurement is consistent with the homogeneous
      prediction within $\sim2\sigma$. Similar positive deviations have also been reported by other
      studies \citep{1972ApJ...172..253S,2004PhRvD..69h3524A}. As noted by
      \citet{2010ApJS..186...94K}, a positive deviation with respect to the homogeneous $\beta=0.6$
      value also suggests that the local Universe could be under-dense on scales of a few hundred
      Mpc. This underdensity could have an impact in explaining the differences between local
      measurements of the expansion rate and those derived from CMB observations
      \citep{2012ApJ...754..131K}.
      
      The values of $\beta$ measured from the mock fractal realizations further support this
      result. Figure \ref{fig:scaling_ndens} shows the mean value and standard deviation of
      the normalized number-counts histograms for the three different values of the fractal
      dimension $D=2.5,\,2.75$ and 2.9, together with the best-fit prediction for the
      2MPZ data. The measured values of $\beta$ are
      \begin{align}\nonumber
        &\beta(D=2.50)=0.51\pm0.05,\\\nonumber
        &\beta(D=2.75)=0.55\pm0.03,\\\nonumber
        &\beta(D=2.90)=0.59\pm0.02,
      \end{align}
      in excellent agreement with the prediction $\beta=0.2\,D$. We can see that the
      value measured from the data lies more than 2 standard deviations away from the fractal
      predictions for $D=2.5$ and 2.75.
      
    \subsubsection{Scaling of the angular 2-point correlation function}
    \label{sssec:scaling_wth}
      In order to test the scaling law for the angular correlation function as a function
      of magnitude limit, Eq. (\ref{eq:scaling_wth}), from the 2MPZ sample, we generated
      four subsamples with different maximum $K_s$ magnitude. These were chosen to have
      the same bright limit $K_{s,{\rm min}}=12.0$ and a varying faint limit
      $K_{s,{\rm max}}=12.5,\,13.0,\,13.5$ and 13.9.
      
      For each of these subsamples we compute the angular two-point correlation function
      using the estimator introduced by \citet{1993ApJ...412...64L}:
      \begin{equation}
        w(\theta)=\frac{DD(\theta)-2DR(\theta)+RR(\theta)}{RR(\theta)},
      \end{equation}
      where $DD,\,DR\,{\rm and}\,RR$ are the normalized counts of pairs of objects
      separated by angle $\theta$ for ``data-data'', ``data-random'' and ``random-random''
      pairs respectively. The correlation function was estimated using the
      software presented in \citet{2012arXiv1210.1833A}, with random catalogues containing 10
      times as many objects as our data. In order to compute the statistical uncertainties in
      $w(\theta)$ we divided our catalogue into $N_{\rm S}=50$ samples, each covering
      approximately the same area. These were chosen evenly in the north and south Galactic
      hemispheres in the region with $|b|>13^\circ$. The covariance matrix $C_{i,j}\equiv
      \langle\Delta w(\theta_i)\Delta w(\theta_j)\rangle$ was estimated as the sample covariance
      matrix of the measurements of $w(\theta)$ made in each of these samples
      scaled by a factor $1/N_{\rm S}$ to account for the smaller area covered by each of
      them:
      \begin{equation}
        \hat{C}_{ij}=\sum_{n=1}^{N_{\rm S}}\frac{[w_{n}(\theta_i)-\bar{w}(\theta_i)]
                     [w_{n}(\theta_j)-\bar{w}(\theta_j)]}{N_{\rm S}(N_{\rm S}-1)},
      \end{equation}
      where $w_n$ is the angular correlation function measured in the $n-$th
      region and $\bar{w}\equiv \sum_n w_n/N_{\rm S}$. The inverse covariance matrix
      used later to compute the $\chi^2$ was estimated in terms of the inverse of
      $\hat{C}_{ij}$ as \citep{2007A&A...464..399H}
      \begin{equation}
        [\tilde{C}^{-1}]_{ij}=\frac{N_S-n_\theta-2}{N_S-1}[\hat{C}^{-1}]_{ij},
      \end{equation}
      where $n_\theta$ is the number of bins of $\theta$ in which the correlation
      function was measured.
      
      As is now well known in the field \citep{2005ApJ...619..147M,2011MNRAS.414..329C},
      the measurements of the angular correlation function show large correlations between
      different angles, and therefore it is important to consider the non-diagonal elements of the
      covariance matrix in order to assess the goodness of fit of a particular model. These
      large off-diagonal elements strongly penalize any deviation from the theoretical model and
      can cause a seemingly valid model to yield a bad minimum $\chi^2$. Furthermore, these
      correlations often cause the covariance matrix to be almost degenerate and difficult to
      invert due to statistical noise. Since the number of independent realizations needed to
      estimate an invertible $n_\theta\times n_\theta$ covariance matrix is usually of the order
      $n_\theta^2$, we performed the fits described below in a reduced range of scales (clearly
      stated in each case).
      
      \begin{figure}
        \centering
        \includegraphics[width=0.49\textwidth]{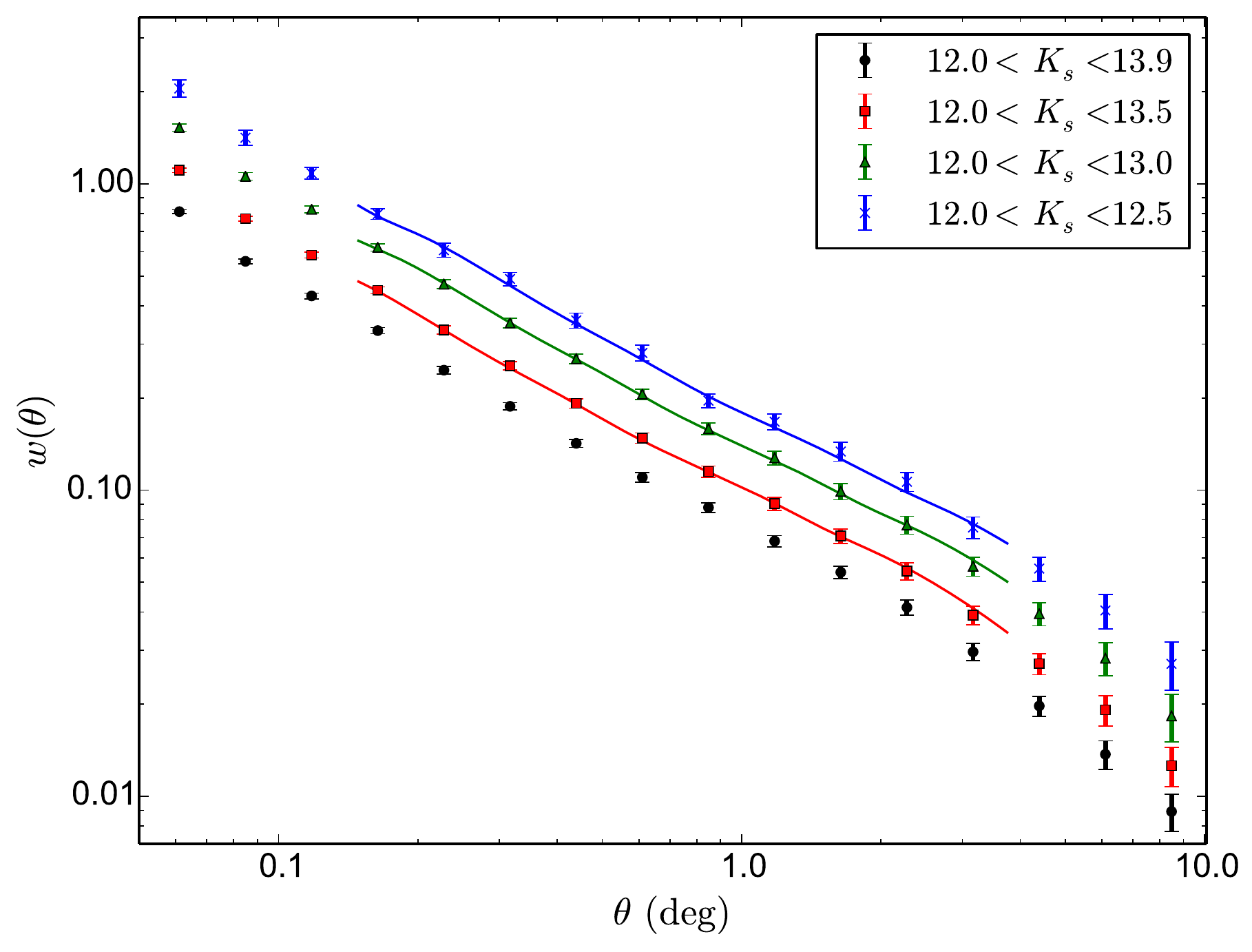}
        \caption{Angular correlation functions for the four $K_s$-magnitude bins considered
        in this analysis. The points with error bars show the actual measurements and estimated
        uncertainties, while the solid lines correspond to the best fit templates rescaled using
        Equation \ref{eq:model2}.}
        \label{fig:scaling_wth}
      \end{figure}
      \begin{table}
        \begin{center}
          \begin{tabular}{l|c|c|c}
            $(K_{s,{\rm min}},K_{s,{\rm max}})$ & $B$ & $\chi^2/{\rm dof}$ & $p$-value \\
            \hline
            (12.0,13.5) & $0.85\pm0.01$ & 0.91 & 0.51 \\
            (12.0,13.0) & $0.71\pm0.01$ & 0.72 & 0.70 \\
            (12.0,12.5) & $0.61\pm0.01$ & 1.45 & 0.16 \\
            \hline
          \end{tabular}
        \end{center}
        \caption{Values of the scaling parameter $B$ measured from the angular correlation function
                 in different magnitude bins, together with their reduced $\chi^2$ and associated 
                 $p$-value for the best-fit model in Eq. \eqref{eq:model2}.}
        \label{tab:scaling}
      \end{table}   
      In order to test the scaling relation in Equation \ref{eq:scaling_wth} we use the 
      measured correlation function in the fiducial magnitude bin $12.0<K_s<13.9$ as a template 
      that we can rescale to fit for the scaling parameter $B$ in the other bins. Other
      possibilities would be using a template based on the theoretical prediction for the
      correlation function or a simple power-law model. The latter option has been shown to be a
      bad approximation to the correlation function on small-scales \citep{2005ApJ...619..147M},
      while the former is affected by theoretical uncertainties. Since our method uses only
      observed quantities, we avoid any of these potential biases, and purely test the
      predicted scaling relation. Our method is therefore:
      \begin{enumerate}
        \item We find the cubic spline $W_{\rm sp}(\theta)$ that interpolates through the
              measurements of the correlation function in our widest magnitude bin.
        \item We use a rescaling of this spline as a template to fit for the scaling parameter
              $B$ in the other three bins. Thus, for each bin we find the parameter $B$ that
              best fits the data given the model
              \begin{equation}\label{eq:model2}
                w(\theta,B)=\frac{W_{\rm sp}(B\theta)}{B}.
              \end{equation}
              This fit is performed by minimizing the corresponding $\chi^2$ using the full
              covariance matrix.
      \end{enumerate}
      In order to avoid using an incorrect template due to the statistical uncertainties in
      the correlation function of the first magnitude bin, caused by shot noise (mainly on
      the smallest scales) and cosmic variance (largest scales), we performed this fit in 
      the range $0^\circ.16<\theta<3^\circ.15$, using 10 logarithmic bins of $\theta$.
      
      The values of $B$ estimated using this method are shown in Table \ref{tab:scaling}
      together with their corresponding $\chi^2$. These results are also displayed in Figure
      \ref{fig:scaling_wth}. In all cases we find that the scaling
      law is a good description of the relation between the correlation functions with
      different magnitude limits. The agreement between the measured values of
      $B$ and the prediction $B\sim\sqrt{F_1/F_2}$ is difficult to address due to the presence
      of a low magnitude cut. However, in the case of the first magnitude bin
      ($B=0.85\pm0.01$), which is less affected by this cut, the agreement is excellent
      ($\sqrt{F_1/F_2}\simeq0.83$). For the narrowest bin ($12.0<K_s<12.5$) we find a slightly
      higher $\chi^2/{\rm d.o.f.}$, which is still statistically insignificant well
      within $2\sigma$.

      \begin{figure}
        \centering
        \includegraphics[width=0.49\textwidth]{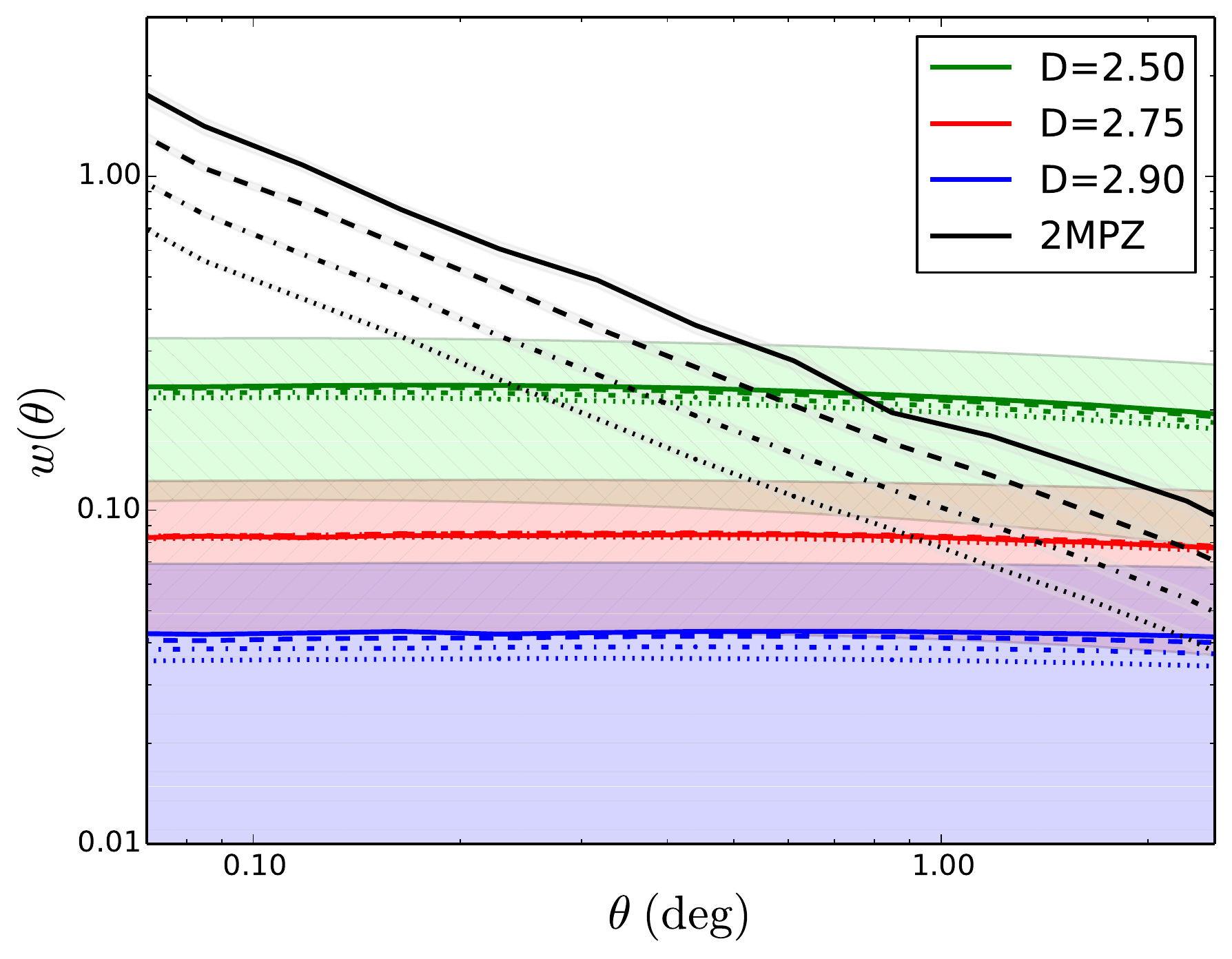}
        \caption{Mean (lines) and variance (coloured bands) of the angular correlation function
        in a suite of 100 mock fractal realizations described in Appendix \ref{app:mocks_fractal}.
        The results for fractal dimensions $D=2.5,\,2.75$ and 2.9 are shown in green, red and
        blue respectively, while the correlation functions measured from 2MPZ are shown as black
        lines. In all cases solid, dashed, dot-dashed and dotted lines correspond
        to the mean correlation function in the magnitude bins $K_s\in(12.0,12.5)$,
        $K_s\in(12.0,13.0)$, $K_s\in(12.0,13.5)$ and $K_s\in(12.0,13.9)$ respectively, and the
        coloured bands around them show the $1\sigma$ deviations. Fractal models show a much
        milder variation in the amplitude of $w(\theta)$ as a function of magnitude limit.}
        \label{fig:scaling_wth_fractals}
      \end{figure}      
      
      As described in Section \ref{ssec:th_scaling}, larger structures
      are found in a perfect fractal model as the survey depth is increased, and thus
      the amplitude of the angular correlation function should stay almost constant
      with the magnitude limit. We have tested this explicitly using the mock fractal
      realizations described in Appendix \ref{app:mocks_fractal}.
      We computed the angular correlation functions, in the same magnitude bins, from a
      suite of 100 mock catalogues with fractal dimensions
      $D=2.5,\,2.75\,{\rm and}\,2.9$. The coloured lines in Figure
      \ref{fig:scaling_wth_fractals} show the mean correlation functions measured for 12
      different cases: $D=2.5$ in green, $D=2.75$ in red and $D=2.9$ in blue, with
      different line styles showing the results for the different magnitude bins (see
      description in caption). The coloured bands around these lines show the $1\sigma$
      dispersion around this mean estimated from the 100 realizations. For comparison,
      the correlation functions measured in 2MPZ are also shown as black lines in this
      plot. Although the variance associated with fractal models is significantly larger
      than in the standard cosmological model, it is easy to see that the amplitude of
      the correlation function varies a lot less with the magnitude limit in these models.
      While this amplitude roughly doubles in the 2MPZ data from the deepest to the
      shallowest magnitude bin, we do not observe variations larger than $\sim10\%$
      between the mean correlation functions of the fractal realizations.
      Although this is further evidence of the compatibility of our our measurements
      with the standard cosmological model, the large variance of the correlation
      function in fractal scenarios makes it difficult to impose tighter constraints on
      them.
  
  \subsection{The angular homogeneity index}\label{ssec:htheta}
    We have studied the possible fractal nature of the galaxy distribution in the 2MPZ
    sample further by analysing the angular homogeneity index $H_2(\theta)$, described in Section
    \ref{ssec:th_htheta}. In order to optimize the use of our data, we have used the estimator
    {\bf E3} described in \citet{2014MNRAS.440...10A}, to measure this quantity. This
    estimator is based on the method used by \citet{2012MNRAS.425..116S} to measure the fractal
    dimension, and makes use of a random catalogue with the same angular mask as the data to 
    correct for edge effects. The process is as follows:
    \begin{enumerate}
      \item For the $i$-th object in the data, we compute $n^d_i(<\theta)$ and $n^r_i(<\theta)$,
            the number of data and random objects respectively found in a spherical cap of radius
            $\theta$ centered on $i$.
      \item For $N_c$ objects in the data thus used as centres of spherical caps, we define the
            scaled counts-in-caps $\mathcal{N}(\theta)$ as
            \begin{equation}
              \mathcal{N}(\theta)=\frac{1}{N_c}\sum_{i=1}^{N_c}
              \frac{n_i^d(<\theta)}{f_r\,n_i^r(<\theta)},
            \end{equation}
            where $f_r\equiv D/R$ is the ratio between the number of data and random objects.
      \item $\mathcal{N}(\theta)$ is directly related to the angular correlation integral
            $G_2(\theta)$ as
            \begin{equation}
              G_2(\theta)=\bar{N}(\theta)\,\mathcal{N}(\theta)-1,
            \end{equation}
            where $\bar{N}(\theta)$ is the expected number of galaxies in a spherical cap of
            radius $\theta$, $\bar{N}(\theta)\equiv \bar{n}_\Omega V(\theta)\equiv
            \bar{n}_\Omega\,2\pi(1-\cos\theta)$, and we have explicitly subtracted the Poisson
            contribution due to linear shot-noise.
      \item The homogeneity index is then estimated by numerical differentiation of $G_2$:
            \begin{equation}
              H_2(\theta)=\frac{d\log G_2(\theta)}{d\log V(\theta)}.
            \end{equation}
    \end{enumerate}
     
    The main advantage of this estimator is that, since it attempts to correct for
    edge-effects using random realizations, it is possible, in principle, to use all 
    the objects in the data as centres for spherical caps of any scale $\theta$, thus 
    minimizing the statistical uncertainties on $H_2$. However, by doing this we can
    potentially bias our estimate of $H_2(\theta)$ towards the homogeneous value $H_2=1$.
    The reason for this is that weighting by the random number counts $n_i^r$ is equivalent
    to assuming that our sample is homogeneous in the parts of the spherical caps that lie
    inside the masked regions. Therefore this potential bias will be more important for
    larger $\theta$, for which a larger fraction of the spherical caps will be masked. In
    order to limit the effects of this bias we have performed two different tests:
    \begin{itemize}
      \item Using the mask described in Section \ref{ssec:sample} (see bottom panel in
            Figure \ref{fig:maps1}) we found, for every unmasked pixel, all other pixels
            lying within a distance $\theta$ from it and computed the fraction
            of those pixels that are unmasked. We thus estimated the average completeness
            of spherical caps as a function of their radius $\theta$. The potential systematic
            edge effects mentioned in the previous paragraph should become more important as this
            completeness decreases, and thus we can limit their impact by constraining our
            analysis to scales with a mean completeness above a given threshold. We determined
            that using a fiducial completeness threshold of $>75\%$ limits the spherical caps
            that can be used for our analysis to scales smaller than $\theta_{\rm max}=40^\circ$.
            A more stringent completeness cut of $85\%$ would translate into a scale cut
            $\theta\lesssim20^\circ$.
      \item For scales below the threshold $\theta_{\rm max}$ it would be desirable to estimate
            the magnitude of the systematic error induced on the measurement of $H_2(\theta)$ by
            the incomplete sky coverage. This can be done using the fact that these systematic
            effects should vanish entirely in the absence of an angular mask, and thus they can
            be quantified by comparing the measurements made on simulated catalogues with and
            without mask. We computed the mean value of $H_2(\theta)$ in our fractal and
            lognormal realizations in these two cases and estimated the average fractional
            deviations between them. For our lognormal realizations, as well as for the fractal
            mocks with $D=2.9$ and $D=2.75$, a small systematic bias smaller than $0.1-0.2\%$
            is found, which increases to $\sim0.6\%$ for $D=2.5$. As we will see below, this
            small bias does not affect any of the results found in this work, and can therefore
            be neglected.
    \end{itemize}
    In view of these results we determined that limiting our analysis to scales $\theta<40^\circ$
    should limit the potential bias of our estimator to an acceptable level.
     
    \begin{figure}
      \centering
      \includegraphics[width=0.49\textwidth]{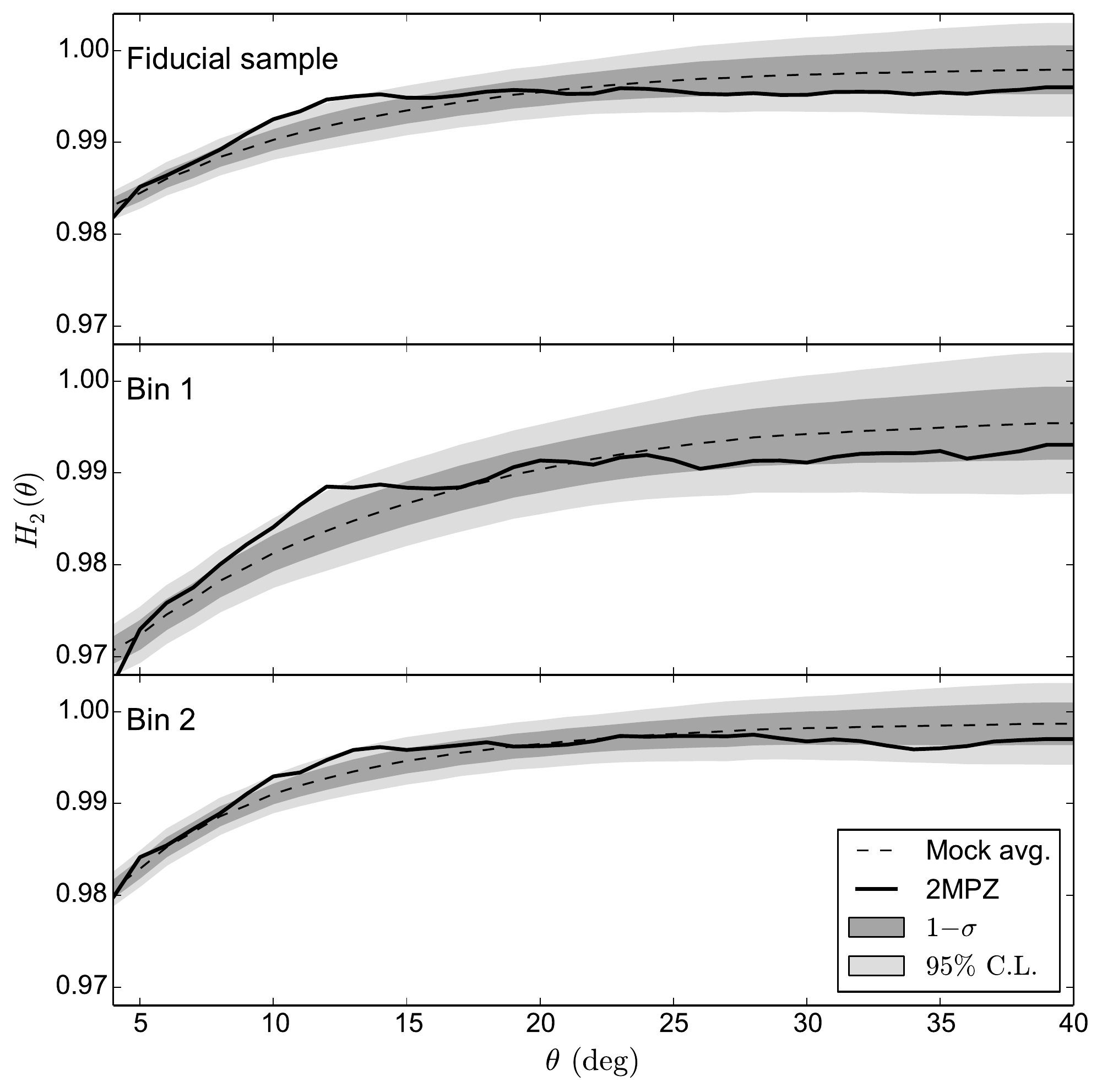}
      \caption{Angular homogeneity index computed in our three samples (see Table \ref{tab:nz_bf}).
      In each panel the solid line shows our measurements from the 2MPZ data, while the dashed,
      thiner lines correspond to the mean value found for a suite of 200 lognormal mock catalogues.
      These mock realizations were used to compute the $1\sigma$ and $1.96\sigma$ (95\% C.L)
      regions, shown as the darker and lighter bands around the measurements.}
      \label{fig:hth_bins}
    \end{figure}
    We computed the homogeneity index for our fiducial sample, as well as for the
    two photometric redshift bins listed in Table \ref{tab:nz_bf}, in order to study the
    evolution of $H_2(\theta)$ with redshift. This was done in the range
    $\theta\in[0^\circ,40^\circ]$, using 40 bins of width $\Delta\theta=1^\circ$. The results are
    shown in Figure \ref{fig:hth_bins}. In descending order each panel presents the results for
    our Fiducial sample, Bin 1 and Bin 2. The solid lines show our measurements from the 2MPZ
    catalogue, while the darker bands around them provide the $1\sigma$ uncertainties, estimated
    as the standard deviation of a suite of 200 lognormal mock catalogues (see Appendix
    \ref{app:mocks_lcdm}). The mean value of the mock realizations, shown as thin dashed lines in
    this figure, can be used as a proxy for the theoretical expectation within $\Lambda$CDM. As
    could be expected, due to the growth of structure and to projection effects (the same angular
    separation corresponds to shorter physical distances on small redshifts), the low-redshift
    sample (Bin 1) is more inhomogeneous than the high-redshift one. It is also worth noting that
    the measurements agree qualitatively well with the mean value of the lognormal realizations.
    This was not guaranteed a priori: even though lognormal realizations are able to reproduce the
    2-point statistics of the galaxy distribution superbly, the homogeneity index depends also on
    higher-order correlations \citep{Bagla:2007tv}.

     \begin{table}
      \begin{center}
        \begin{tabular}{l|c|c|c}
          Sample name & $\theta_H$ (2MPZ) & $\theta_H$ (Mocks) & $f_{\rm above}$ \\
          \hline
          Fiducial & $35^\circ$ & $26^\circ\pm5^\circ$ & $10.5\%$ \\
          Bin 1    & $39^\circ$ & $28^\circ\pm5^\circ$ & $14\%$ \\
          Bin 2    & $24^\circ$ & $24^\circ\pm5^\circ$ & $54.5\%$\\
          \hline
        \end{tabular}
      \end{center}
      \caption{Angular homogeneity scale for our three samples. The second column shows the 
               values of $\theta_H$ measured from the data, while the third column shows the
               mean value and variance estimated from the mock catalogues. The last column
               shows the fraction of mock realizations in which homogeneity is attained on
               scales larger than the value of $\theta_H$ measured in the data.}
      \label{tab:theta_h}
    \end{table}                
    The homogeneity index, as measured in 2MPZ, seems to approach the perfect homogenous
    prediction $H_2=1$ on large scales, although it deviates slightly from it. As we have
    mentioned, these deviations are expected within $\Lambda$CDM, and we should therefore
    quantify their significance. In order to do so we have estimated the angular scale of
    homogeneity $\theta_H$, proposed in \citet{2014MNRAS.440...10A} as the largest angle
    for which the measured value of $H_2$ deviates from the homogeneous value of 1 at 95\% C.L.
    (i.e. $1-H_2(\theta_H)=1.96\,\sigma_{H_2}(\theta_H)$, where $\sigma_{H_2}$ is the
    statistical error on $H_2$). We computed $\theta_H$ for the same three samples as well
    as for their corresponding 200 mock realizations in order to quantify the expected variance
    of this quantity. For each sample we estimated the mean value of $\theta_H$ and its variance
    from the mock lognormal catalogues, as well as the fraction of mock catalogues in which
    homogeneity is reached on scales larger than the value of $\theta_H$ found in the 2MPZ data
    ($f_{\rm above}$). The results are summarized in Table \ref{tab:theta_h}: in agreement with
    our previous result, we find that our samples containing low-redshift objects (``Fiducial''
    and ``Bin 1'') reach homogeneity on scales exceeding the average value found in the mock
    realizations by about 2 standard deviations. The higher-$z$ sample (``Bin 2''), nevertheless,
    agrees very well with the expected value of $\theta_H\simeq24^\circ$. In all cases at least
    $10\%$ of the mock realizations were found to reach homogeneity on scales larger than the
    values of $\theta_H$ measured in the data, which places the level of disagreement with the
    $\Lambda$CDM expectation well below $2\sigma$.
    
    \begin{figure}
      \centering
      \includegraphics[width=0.49\textwidth]{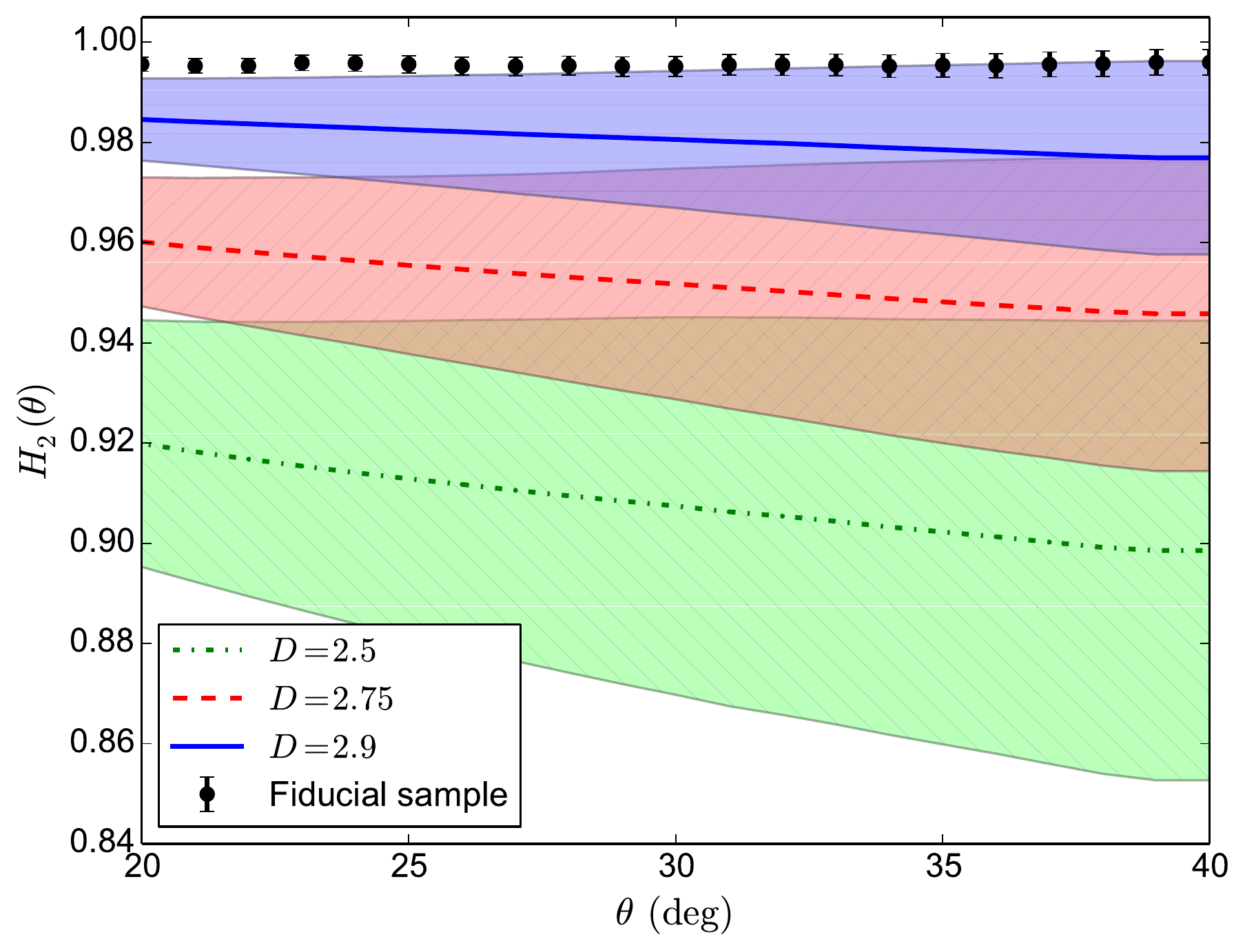}
      \caption{Angular homogeneity index measured from the 2MPZ data (points with error bars),
      as well as the mean value and variance (solid lines and light coloured bands respectively)
      estimated for the fractal $\beta$-model from the suite of mock catalogues described in
      Appendix \ref{app:mocks_fractal}. The results for the fractal models correspond to 
      fractal dimensions $D=2.5$ (bottom curve, red), $D=2.75$ (middle curve, green) and 
      $D=2.9$ (top curve, blue).}
      \label{fig:hth_fractals}
    \end{figure}     
    Thus far we have shown that the measurements of $H_2(\theta)$ are fully compatible with the
    expectations of the standard cosmological model. However, we can also use these
    measurements of the asymptotic value of $H_2(\theta)$ to explore the viability of fractal
    models. As has been noted in the literature \citep{1997EL.....40..491D}, the fractal nature
    of a three-dimensional point distribution cannot be completely determined from its
    distribution projected on the sphere in a model-independent way, and therefore we cannot
    hope to rule out fractal models in general with this test. However, using the fractal
    $\beta$-model described in Appendix \ref{app:mocks_fractal} we can study the values of
    $H_2(\theta)$ that can be expected from a cascading fractal model, which could potentially
    apply (at least qualitatively) to more general scenarios. Figure \ref{fig:hth_fractals} shows
    the mean value and expected variance of $H_2(\theta)$ on large scales ($\theta>20^\circ$)
    computed from a suite of 100 $\beta$-model realizations with fractal dimensions
    $D=2.5,\,2.75$ and $2.9$, together with our measurements for the fiducial 2MPZ sample. The
    large-scale behaviour of the different projected distributions is clearly distinct, and
    allows us to quantify the disagreement between our measurements and these fractal models.
    We did so by computing the number of fractal realizations in which the value of $H_2(\theta)$
    is consistently larger than our measurement from 2MPZ in the last 10 angular bins (i.e.
    $\theta>30^\circ$), which we can interpret as the probability of finding a fractal Universe
    that is at least as isotropic as ours as implied by our measurements. While 12 of our 100
    simulations with $D=2.9$ are found above the measurements in our fiducial sample, this is
    not the case in any of the realizations with $D=2.75$ or 2.5. This result reinforces the
    compatibility of our measurements with the standard cosmological model.

  \subsection{Systematics}\label{ssec:systematics}
    In this Section we present a set of tests aimed at discarding possible
    systematic effects in our results which could arise from the sample selection described
    in Section \ref{ssec:sample}.

    \subsubsection{Clustering systematics}\label{sssec:sys_clustering}
    \begin{figure}
      \centering
      \includegraphics[width=0.45\textwidth]{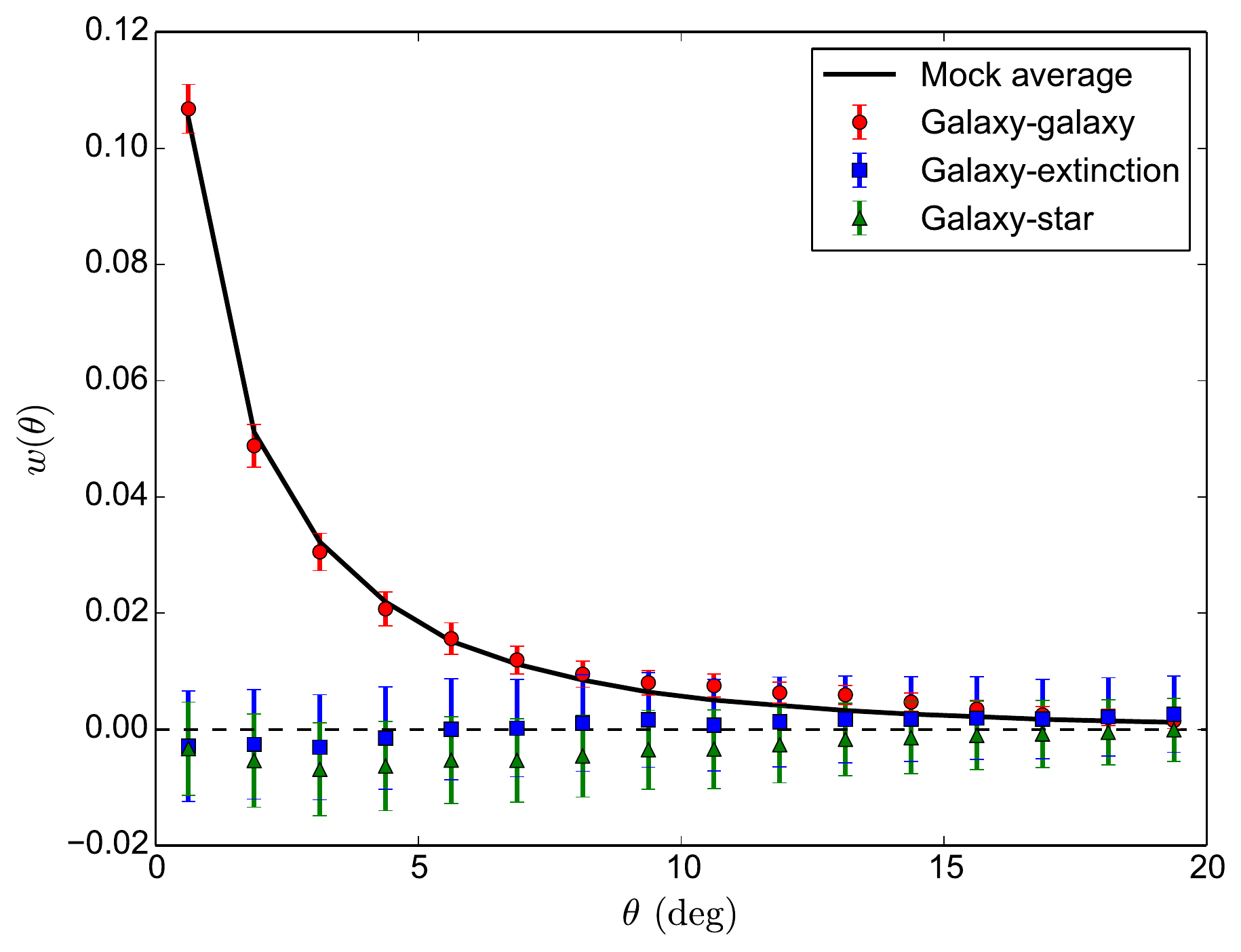}
      \includegraphics[width=0.45\textwidth]{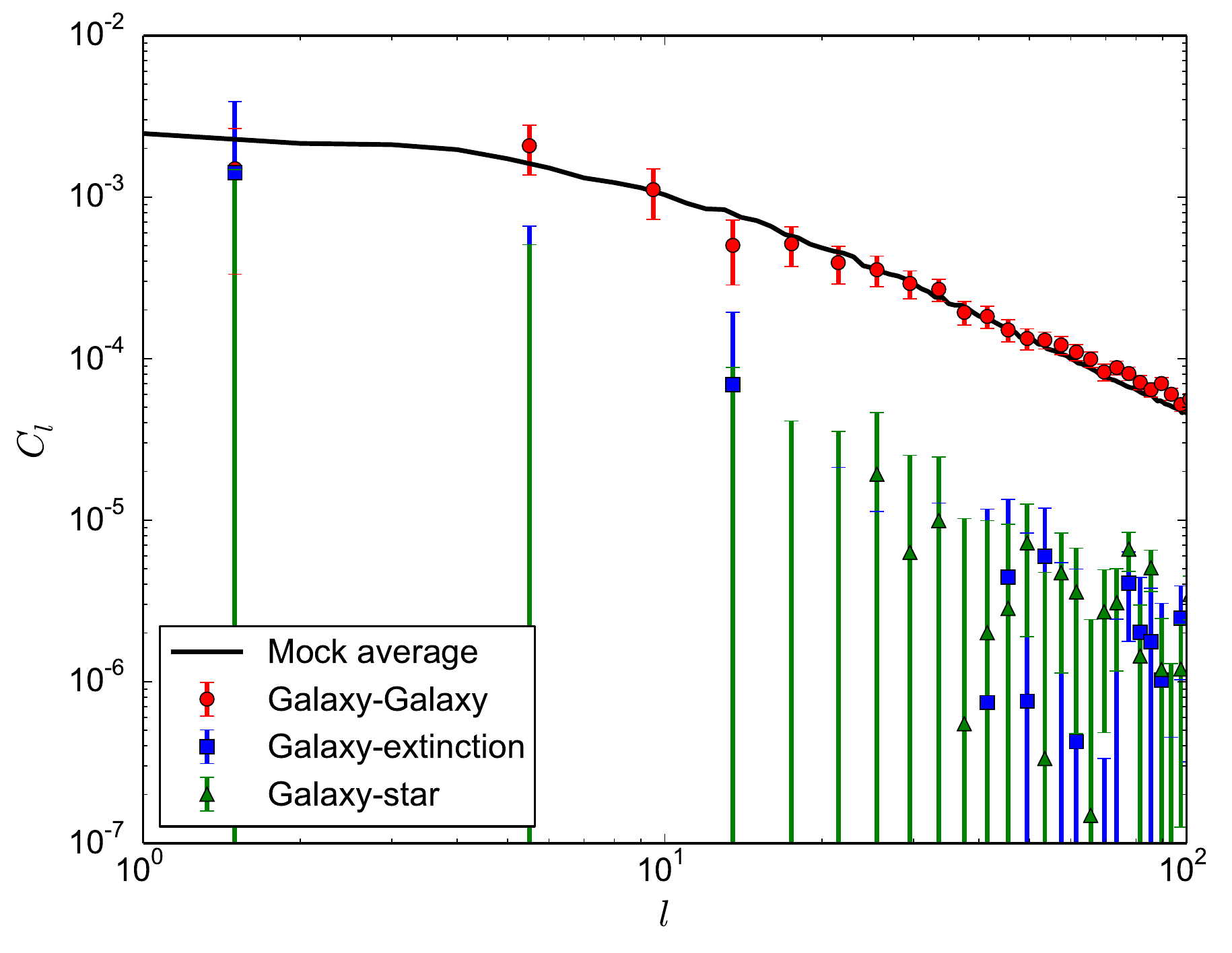}
      \caption{Auto- and cross- correlation in real space (top panel) and harmonic space
               (bottom panel) of the galaxy density field for our fiducial sample with
               the two potential sources of systematics: dust extinction (blue) and
               star density (green). The red points show the galaxy auto-correlation, which
               has a much bigger amplitude than any of the cross-correlations, both of which
               are consistent with 0. The correlations were computed for our fiducial mask,
               and the error bars were estimated using 100 mock catalogues
               described in Appendix \ref{app:mocks_lcdm}.}
      \label{fig:mask_corr}
    \end{figure}
    In order to ensure that the cuts in dust extinction and star density specified in
    Section \ref{ssec:sample} are enough to prevent any significant
    systematic effect in the 2-point clustering statistics of the galaxy density field,
    we studied its cross-correlation with these two possible systematics.
    
    We first generated maps of the anisotropies in the 2MPZ galaxy density, $A_K$ and
    $n_{\rm star}$. This was done by computing, for each observable $x$ and in each pixel $i$,
    the quantity $\delta^i_x=(x^i-\bar{x})/\bar{x}$, where $\bar{x}$ is the mean value of
    $x$ averaged over all unmasked pixels. We then computed the cross- and auto-correlations
    of each pair of observables, $w_{x,y}(\theta)=\langle\delta_x\delta_y\rangle$, where the
    expectation value was estimated by averaging over all pairs of unmasked pixels subtending
    an angle $\theta$. Figure \ref{fig:mask_corr} shows these cross correlations for our fiducial
    sample (see Section \ref{ssec:bias}). The errors on these measurements were computed
    as the standard deviation for a suite of 100 mock galaxy catalogues described in
    Appendix \ref{app:mocks_lcdm}. In both cases the cross correlation of the galaxy 
    overdensity with each systematic is compatible with 0, thus confirming our choice of
    $A_{K,{\rm max}}$ and $n_{\rm star, max}$. The bottom panel of Figure \ref{fig:mask_corr} shows
    the same auto- and cross- correlations in harmonic space (i.e. the power spectrum
    $C_l$), which confirm this result. Since, as stated above, the main contribution to the
    homogeneity index $H_2$ is due to the two-point correlation function, we do not expect
    any significant systematic effect on this quantity either.

  \subsubsection{Hemispherical differences}\label{sssec:sys_hemisph}

    Assuming an isotropic galaxy distribution, an incorrect choice of $A_K$ or $n_{\rm star,max}$
    could cause an asymmetry in the properties of the galaxy sample in the north and south
    Galactic hemispheres. Likewise, errors in the calibration of the two twin facilities used to
    compile 2MASS, located in the two terrestrial hemispheres, could potentially generate a similar
    asymmetry with respect to the equatorial plane.  Thus, investigating 
    the presence of hemispherical differences is a good way of identifying systematic effects
    in a full-sky galaxy survey. Furthermore, this type of effects have been studied in
    different cosmological observations. Probably the most notable of these is the CMB dipolar
    asymmetry detected at $\sim3.5\sigma$ in both Planck and WMAP \citep{2004MNRAS.349..313P,
    2007ApJ...660L..81E,2009ApJ...699..985H,2014A&A...571A..23P,2014ApJ...784L..42A} 
    (although see \citep{2015JCAP...01..008Q}). Similar
    studies have been conducted with other datasets, including 2MASS \citep{2012MNRAS.427.1994G,
    2014JCAP...10..070A,2014MNRAS.445L..60Y}, radio galaxies \citep{2014MNRAS.441.2392F}, luminous
    red galaxies \citep{2010JCAP...05..027P}, radio sources \citep{2014MNRAS.441.2392F} and
    high-redshift quasars \citep{2009JCAP...09..011H}, finding however only marginal dipolar
    signals, if any. Understanding the origin of these asymmetries, when they arise, can not
    only shed light on the possible systematic effect affecting CMB measurements, but also tell
    us something about our relative motion with respect to the CMB rest frame
    \citep{2010PhRvD..82d3530I,2011ApJ...741...31B}. Therefore, investigating the presence of
    these asymmetries in different datasets is also interesting in its own right, besides it
    being an additional test of the isotropy of the galaxy distribution.

    We have searched for hemispherical differences both in the overall galaxy number counts and
    in the clustering variance of the galaxy overdensity.

    \paragraph*{Number counts.} We have studied the differences in the number of galaxies observed
    in two opposite hemispheres in our fiducial dataset ($12<K<13.9$) in relation to the variance
    of this difference expected within the standard cosmological model. In order to do so we
    considered hemispheres defined in terms of planes tilted by an angle $\alpha$ with respect to
    three fundamental planes: Galactic, ecliptic and equatorial. The reason for considering
    a varying angle $\alpha$ is that any potential systematic effects (for example, instrumental
    differences in the case of the equatorial plane) would become gradually more evident as
    $\alpha\rightarrow0$. For each pair of hemispheres we computed the angular number density of
    galaxies in each of them as $\bar{n}_{\Omega}=N_{\rm gal}/(4\pi\,f_{\rm sky})$, where
    $N_{\rm gal}$ is the number of galaxies observed in that hemisphere and $f_{\rm sky}$ is the
    corresponding observed sky fraction (note that due to the incomplete sky coverage,
    $f_{\rm sky}$ will be different for both hemispheres and also for different values of
    $\alpha$). In terms of these measurements, as a statistical observable  we used the relative
    difference between both number densities:
    \begin{equation}
      \Delta_n\equiv
         \frac{2\,|\bar{n}^N_\Omega-\bar{n}^S_\Omega|}{\bar{n}^N_\Omega+\bar{n}^S_\Omega}.
    \end{equation}

    \begin{figure}
      \centering
      \includegraphics[width=0.49\textwidth]{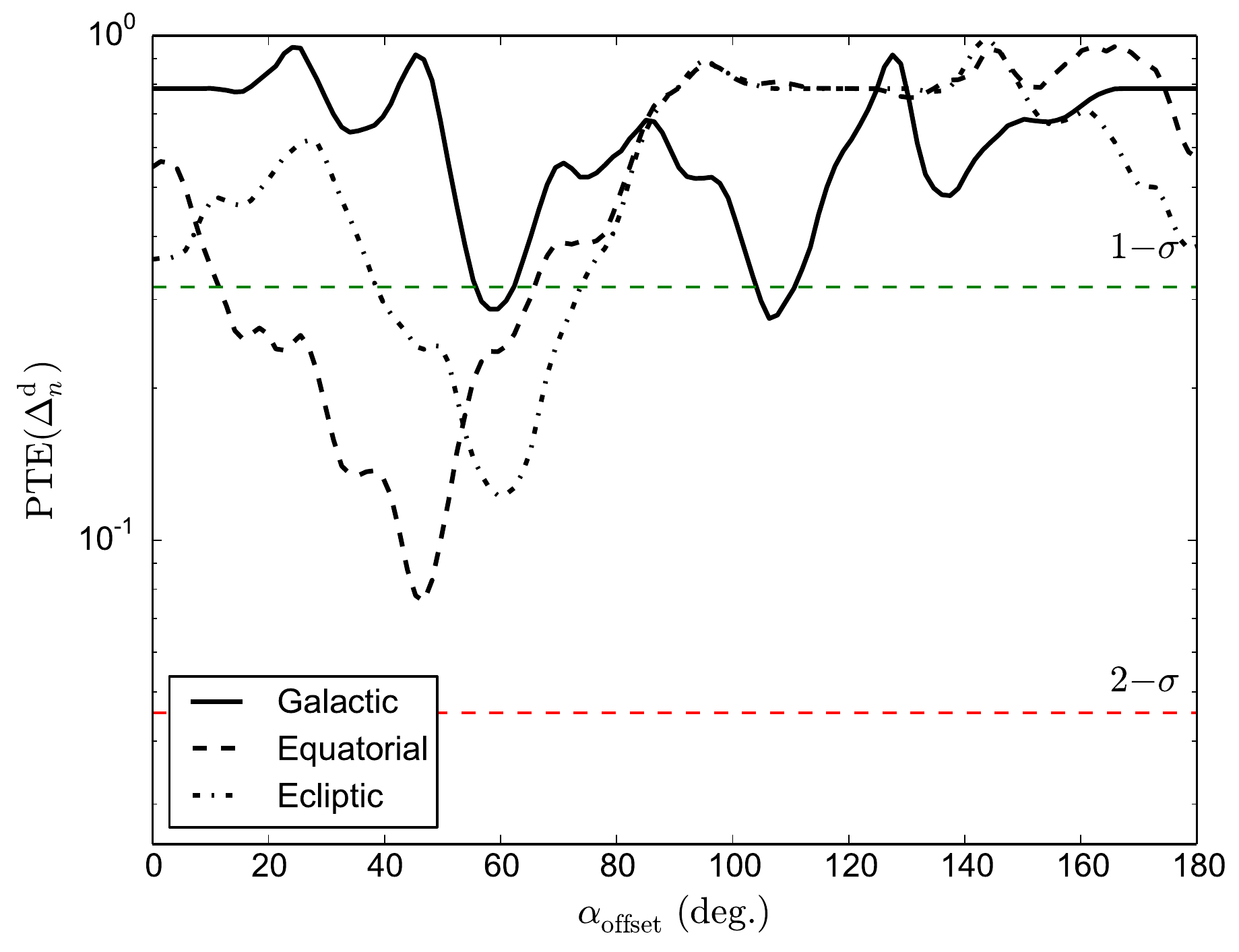}
      \caption{Probability to exceed for the normalized difference in the galaxy number counts
               measured in hemispheres defined by an angle $\alpha$ with respect to the
               three fundamental planes: Galactic (solid line), equatorial (dashed line) and
               ecliptic (dash-dotted line). In all cases the observed asymmetry can be
               explained by the statistical uncertainties for a $\Lambda$CDM model well within
               $2\sigma$.}
      \label{fig:hemisph_p_ndens}
    \end{figure}
    \begin{figure}
      \centering
      \includegraphics[width=0.49\textwidth]{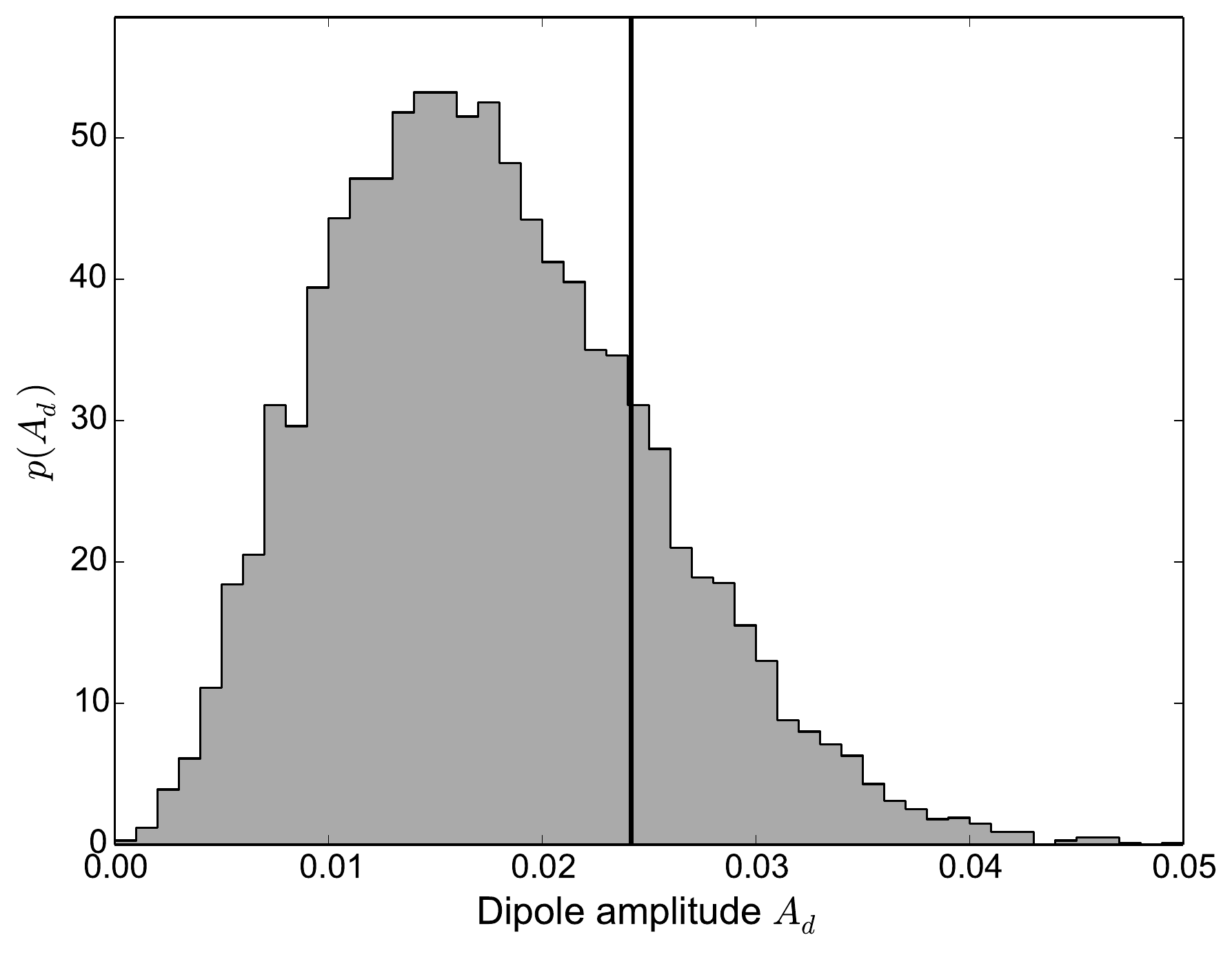}
      \caption{Probability distribution for the amplitude of the dipole in the local variance maps
               for disks of size $10^\circ$, derived from a suite of 10000 lognormal mock
               realizations. The value of $A_d$ measured in our fiducial sample is shown as a
               vertical solid line in this Figure, and is in good agreement (within $1.3\sigma$)
               with the expected $\Lambda$CDM value.}
      \label{fig:hemisph_p_chi2}
    \end{figure}

    We computed the value of $\Delta_n$ in the data as well as in 1000 independent mock
    catalogues, and used the mock measurements to estimate the probability distribution
    of this observable $p(\Delta_n|\alpha)$. We then characterized the compatibility between the
    number densities in each hemisphere measured in the data by computing the fraction of mocks
    for which we find a value of $\Delta_n$ larger than the one measured in the data,
    $\Delta_n^d$:
    \begin{equation}
      {\rm PTE}(\Delta_n^{\rm d},\alpha)\equiv\int_{\Delta_n^{\rm d}}^\infty
      p(\Delta_n|\alpha)\,d\Delta_n.
    \end{equation}
    We note that, even for the fairly large volume probed by 2MASS ($\chi(z_{\rm max})\sim
    800\,{\rm Mpc}/h$), the variance of $\Delta_n$ is by far dominated by clustering variance,
    and not Poisson noise. In comparison with the total error (cosmic variance + Poisson),
    computed from our lognormal realizations, Poisson errors are a factor of 12-13 times
    smaller. This must be taken into account when interpreting the significance
    of the observed asymmetries. The values of ${\rm PTE}(\Delta^d_n,\alpha)$ estimated through
    this method are shown in Figure \ref{fig:hemisph_p_ndens} for the three fundamental planes.
    In no case do we find evidence for a hemispherical asymmetry in the galaxy number counts
    with a significance larger than $2\sigma$.
    
    \paragraph*{Clustering.}
      In order to detect possible hemispherical power asymmetries in our data we have performed
      an analysis similar to that carried out recently by \citet{2014ApJ...784L..42A} on CMB data.
      The method proceeds as follows:
      \begin{enumerate}
        \item We first find disks of different angular sizes centered on the pixels of a HEALPix
              map of resolution ${\tt Nside}=16$ (i.e. 3072 pixels). For this analysis we
              considered disks of radius $\theta=10^\circ$ and $\theta=20^\circ$.
        \item We create a map of the overdensity field for our fiducial sample as described in
              Section \ref{ssec:bias} using our fiducial pixel resolution ${\tt Nside}=64$. From
              this map we compute the variance of the overdensity field inside each of the disks
              found in the previous step. This yields a low-resolution map of the local variance.
              We neglect any disks for which more than $90\%$ of the pixels were masked.
        \item We repeat the previous step on a suite of 10000 mock lognormal realizations. From
              these we compute the mean local variance across the sky, as well as its standard
              deviation.
        \item We subtract this mean map from the local variance map computed for our data, and
              fit a dipole to the resulting zero-mean map using an inverse-variance scheme. We
              perform the same operation on the local variance maps found for the 10000 mocks
              and store the values of the dipole amplitude $A_d$ found in each case.
        \item We characterize the significance of the power asymmetry in our data in terms
              of the number of mocks found with a value of $A_d$ larger than the one found in
              our sample.
      \end{enumerate}
      Figure \ref{fig:hemisph_p_chi2} shows the distribution of dipole amplitudes $A_d$ computed
      from the 10000 mock catalogues, for disks of size $10^\circ$, together with the value found
      in the data. The quantitative results are displayed in Table \ref{tab:p_chi2}. For disks of
      both $10^\circ$ and $20^\circ$ we consistently found a dipole of amplitude $A_d\sim0.025$ in
      the direction $(l,b)\sim(310^\circ,5^\circ)$, which is close to the direction of the dipole
      found by \citet{2012MNRAS.427.1994G} using 2MASS. The amplitude of this dipole is, however,
      in excellent agreement (within $\sim1.5\sigma$) with the variance expected within
      $\Lambda$CDM. We can thus conclude that there are no significant hemispherical asymmetries
      in our data.
     
     \begin{table}
        \begin{center}
          \begin{tabular}{l|c|c|c}
            Disk radius & $A_d$ & $(l,b)$ & $p-$value\\
            \hline
            $10^\circ$   & $0.024$ & $(311^\circ,0^\circ)$ & $0.19\,\,\,(1.3\sigma)$ \\
            $20^\circ$   & $0.028$ & $(320^\circ,6^\circ)$ & $0.13\,\,\,(1.5\sigma)$ \\
            \hline
          \end{tabular}
        \end{center}
        \caption{Values dipole amplitude and direction in the local variance maps constructed
                 for our fiducial sample, for disks of aperture $10^\circ$ and $20^\circ$. The
                 fourth column shows the fraction of mock catalogues found with a dipole amplitude
                 $A_d$ larger than the one measured in the data. In both cases the data is
                 found to be compatible with the expected $\Lambda$CDM dipole well within
                 $2\sigma$.}
        \label{tab:p_chi2}
      \end{table}     
     
\section{Discussion}\label{sec:discussion}
  In this paper we have presented an analysis of the homogeneity and isotropy of the low-redshift
  galaxy distribution using data from 2MPZ. Making use of a set of observational probes relying
  mainly on angular information we have been able to study possible deviations of a fractal nature
  from the standard cosmological model. These probes included the scaling laws of the source number
  counts and angular correlation function as a function of magnitude limit and the measurement of
  the so-called homogeneity scale, which can be interpreted as the effective correlation dimension
  of the projected distribution. The 2MPZ sample is particularly well suited for this kind of
  studies for several reasons: first of all, its almost complete sky coverage makes it possible to
  study clustering on the largest angular scales. Secondly, the low median redshift of the survey
  implies that we can probe the evolution of the galaxy distribution at late times, where
  deviations from statistical homogeneity and isotropy due to gravitational clustering are
  potentially larger. Finally, the availability of photometric redshifts for all 2MPZ sources makes
  it possible to explore the evolution of the homogeneity index with redshift and compare that
  evolution with the expected behaviour within the $\Lambda$CDM model. We have found that, in
  terms of these observables, the data is in excellent agreement with the standard cosmological
  model, and no significant departure from its predictions has been observed.
  
  The main results of this paper can be summarized as follows:
  \begin{itemize}
    \item Our measurements of the homogeneity index $H_2(\theta)$ show a good agreement of the
          data with the standard cosmological model, and we have verified that the galaxy
          distribution approaches homogeneity within the expected range of angular scales. We have
          shown that this agreement holds also as a function of cosmic time by repeating the
          analysis in two bins of photometric redshift. We repeated this analysis on a suite of
          mock fractal realizations and found that none of those with fractal dimensions
          $D\lesssim2.75$ approached homogeneity faster than the 2MPZ sample.
    \item We have shown that the measurements of the scaling laws for number counts and correlation
          function closely follow the expectation of a statistically homogeneous cosmology, while
          our tests using a particular fractal scenario (see Appendix \ref{app:mocks_fractal}) have
          shown that these models display evident tension with these observables for fractal
          dimensions $D\lesssim2.75$.
    \item As part of our search for systematics we perform an extra test of statistical isotropy
          by investigating the presence of hemispherical asymmetries in our data. We find a
          dipole in the clustering variance of the data in the same direction ($(l,b)\sim
          (310^\circ,5^\circ)$) as previous studies. The amplitude of this dipole is,
          nevertheless, in perfect agreement with the variance expected within $\Lambda$CDM.
  \end{itemize}

  Testing the validity of the CP is a necessary step before using any cosmological probe that,
  implicitly or explicitly assumes this validity. The observational evidence backing the
  large-scale homogeneity and isotropy of the matter distribution has grown significantly
  in the last few decades, and our results certainly support this evidence in the local
  Universe. In the near future it will be possible to impose further constraints on possible
  departures from the CP by performing this kind of analyses on deeper wide-area surveys. We
  plan to apply our methodology to the forthcoming WISE-based photometric catalogues probing
  75\% of the sky at redshifts $z<0.5$
  \citep[Bilicki et al.\ 2015, in prep.]{2014arXiv1408.0799B}, as well as to the Dark Energy
  Survey data \citep{Flaugher:2004vg} covering less of the sky but at larger depths.

\section*{Acknowledgments}
  We would like to thank Pedro Ferreira, Sigurd N\ae{}ss and John Peacock for useful comments
  and discussions, as well as the anonymous journal referee for their insightful comments. DA
  is supported by ERC grant 259505. MB acknowledges the financial assistance of the South
  African National Research Foundation (NRF), as well as of the Polish National Science Centre
  under contract \#UMO-2012/07/D/ST9/02785. We acknowledge the effort made by the Wide Field
  Astronomy Unit at the Institute for Astronomy, Edinburgh in archiving the 2MPZ catalogue,
  which can be accessed at \url{http://surveys.roe.ac.uk/ssa/TWOMPZ}.

\setlength{\bibhang}{2.0em}
\setlength\labelwidth{0.0em}
\bibliography{paper}

\begin{thebibliography}{}
 \providecommand{\href}[2]{#2}
  \providecommand{\doi}[1]{\href{http://dx.doi.org/#1}{doi:#1}}
  \providecommand{\eprint}[1]{\href{http://arxiv.org/abs/#1}{arXiv:#1}}

\bibitem[\protect\citeauthoryear{{Afshordi}, {Loh} \& {Strauss}}{{Afshordi}
  et~al.}{2004}]{2004PhRvD..69h3524A}
{Afshordi} N.,  {Loh} Y.-S.,    {Strauss} M.~A.,  2004, \prd, 69, 083524,
  \eprint{astro-ph/0308260}

\bibitem[\protect\citeauthoryear{{Ahn} et~al.,}{{Ahn} et~al.}{2012}]{SDSS.DR9}
{Ahn} C.~P.  et~al., 2012, \apjs, 203, 21, \eprint{1207.7137}

\bibitem[\protect\citeauthoryear{{Akrami}, {Fantaye}, {Shafieloo}, {Eriksen},
  {Hansen}, {Banday} \& {G{\'o}rski}}{{Akrami}
  et~al.}{2014}]{2014ApJ...784L..42A}
{Akrami} Y.,  {Fantaye} Y.,  {Shafieloo} A.,  {Eriksen} H.~K.,  {Hansen} F.~K.,
   {Banday} A.~J.,    {G{\'o}rski} K.~M.,  2014, \apjl, 784, L42,
  \eprint{1402.0870}

\bibitem[\protect\citeauthoryear{{Alonso}}{{Alonso}}{2012}]{2012arXiv1210.1833A}
{Alonso} D.,  2012, ArXiv e-prints, \eprint{1210.1833}

\bibitem[\protect\citeauthoryear{{Alonso}, {Bueno Belloso}, {S{\'a}nchez},
  {Garc{\'{\i}}a-Bellido} \& {S{\'a}nchez}}{{Alonso}
  et~al.}{2014}]{2014MNRAS.440...10A}
{Alonso} D.,  {Bueno Belloso} A.,  {S{\'a}nchez} F.~J.,
  {Garc{\'{\i}}a-Bellido} J.,    {S{\'a}nchez} E.,  2014, \mnras, 440, 10,
  \eprint{1312.0861}

\bibitem[\protect\citeauthoryear{{Appleby} \& {Shafieloo}}{{Appleby} \&
  {Shafieloo}}{2014}]{2014JCAP...10..070A}
{Appleby} S.,  {Shafieloo} A.,  2014, \jcap, 10, 70, \eprint{1405.4595}

\bibitem[\protect\citeauthoryear{Bagla, Yadav \& Seshadri}{Bagla
  et~al.}{2007}]{Bagla:2007tv}
Bagla J.,  Yadav J.,    Seshadri T.,  2007, Mon.Not.Roy.Astron.Soc., 390, 829,
  \eprint{0712.2905}

\bibitem[\protect\citeauthoryear{{Beutler} et~al.,}{{Beutler}
  et~al.}{2011}]{2011MNRAS.416.3017B}
{Beutler} F.  et~al., 2011, \mnras, 416, 3017, \eprint{1106.3366}

\bibitem[\protect\citeauthoryear{{Bilicki}, {Chodorowski}, {Jarrett} \&
  {Mamon}}{{Bilicki} et~al.}{2011}]{2011ApJ...741...31B}
{Bilicki} M.,  {Chodorowski} M.,  {Jarrett} T.,    {Mamon} G.~A.,  2011, \apj,
  741, 31, \eprint{1102.4356}

\bibitem[\protect\citeauthoryear{{Bilicki}, {Jarrett}, {Peacock}, {Cluver} \&
  {Steward}}{{Bilicki} et~al.}{2014}]{2014ApJS..210....9B}
{Bilicki} M.,  {Jarrett} T.~H.,  {Peacock} J.~A.,  {Cluver} M.~E.,    {Steward}
  L.,  2014, \apjs, 210, 9, \eprint{1311.5246}

\bibitem[\protect\citeauthoryear{{Bilicki}, {Peacock}, {Jarrett}, {Cluver} \&
  {Steward}}{{Bilicki} et~al.}{2014}]{2014arXiv1408.0799B}
{Bilicki} M.,  {Peacock} J.~A.,  {Jarrett} T.~H.,  {Cluver} M.~E.,    {Steward}
  L.,  2014, ArXiv e-prints, \eprint{1408.0799}

\bibitem[\protect\citeauthoryear{{Blake} et~al.,}{{Blake}
  et~al.}{2011}]{2011MNRAS.415.2892B}
{Blake} C.  et~al., 2011, \mnras, 415, 2892, \eprint{1105.2862}

\bibitem[\protect\citeauthoryear{{Castagnoli} \& {Provenzale}}{{Castagnoli} \&
  {Provenzale}}{1991}]{1991A&A...246..634C}
{Castagnoli} C.,  {Provenzale} A.,  1991, \aap, 246, 634

\bibitem[\protect\citeauthoryear{{Chon}, {Challinor}, {Prunet}, {Hivon} \&
  {Szapudi}}{{Chon} et~al.}{2004}]{2004MNRAS.350..914C}
{Chon} G.,  {Challinor} A.,  {Prunet} S.,  {Hivon} E.,    {Szapudi} I.,  2004,
  \mnras, 350, 914, \eprint{astro-ph/0303414}

\bibitem[\protect\citeauthoryear{{Coles} \& {Jones}}{{Coles} \&
  {Jones}}{1991}]{1991MNRAS.248....1C}
{Coles} P.,  {Jones} B.,  1991, \mnras, 248, 1

\bibitem[\protect\citeauthoryear{{Collister} \& {Lahav}}{{Collister} \&
  {Lahav}}{2004}]{2004PASP..116..345C}
{Collister} A.~A.,  {Lahav} O.,  2004, \pasp, 116, 345,
  \eprint{astro-ph/0311058}

\bibitem[\protect\citeauthoryear{{Crocce}, {Cabr{\'e}} \&
  {Gazta{\~n}aga}}{{Crocce} et~al.}{2011}]{2011MNRAS.414..329C}
{Crocce} M.,  {Cabr{\'e}} A.,    {Gazta{\~n}aga} E.,  2011, \mnras, 414, 329,
  \eprint{1004.4640}

\bibitem[\protect\citeauthoryear{{Durrer}}{{Durrer}}{2011}]{2011RSPTA.369.5102D}
{Durrer} R.,  2011, Royal Society of London Philosophical Transactions Series
  A, 369, 5102, \eprint{1103.5331}

\bibitem[\protect\citeauthoryear{{Durrer}, {Eckmann}, {Sylos Labini},
  {Montuori} \& {Pietronero}}{{Durrer} et~al.}{1997}]{1997EL.....40..491D}
{Durrer} R.,  {Eckmann} J.-P.,  {Sylos Labini} F.,  {Montuori} M.,
  {Pietronero} L.,  1997, EPL (Europhysics Letters), 40, 491,
  \eprint{astro-ph/9702116}

\bibitem[\protect\citeauthoryear{{Eriksen}, {Banday}, {G{\'o}rski}, {Hansen} \&
  {Lilje}}{{Eriksen} et~al.}{2007}]{2007ApJ...660L..81E}
{Eriksen} H.~K.,  {Banday} A.~J.,  {G{\'o}rski} K.~M.,  {Hansen} F.~K.,
  {Lilje} P.~B.,  2007, \apjl, 660, L81, \eprint{astro-ph/0701089}

\bibitem[\protect\citeauthoryear{{Fern{\'a}ndez-Cobos}, {Vielva}, {Pietrobon},
  {Balbi}, {Mart{\'{\i}}nez-Gonz{\'a}lez} \& {Barreiro}}{{Fern{\'a}ndez-Cobos}
  et~al.}{2014}]{2014MNRAS.441.2392F}
{Fern{\'a}ndez-Cobos} R.,  {Vielva} P.,  {Pietrobon} D.,  {Balbi} A.,
  {Mart{\'{\i}}nez-Gonz{\'a}lez} E.,    {Barreiro} R.~B.,  2014, \mnras, 441,
  2392, \eprint{1312.0275}

\bibitem[\protect\citeauthoryear{{Fixsen}, {Cheng}, {Gales}, {Mather}, {Shafer}
  \& {Wright}}{{Fixsen} et~al.}{1996}]{1996ApJ...473..576F}
{Fixsen} D.~J.,  {Cheng} E.~S.,  {Gales} J.~M.,  {Mather} J.~C.,  {Shafer}
  R.~A.,    {Wright} E.~L.,  1996, \apj, 473, 576, \eprint{astro-ph/9605054}

\bibitem[\protect\citeauthoryear{Flaugher}{Flaugher}{2005}]{Flaugher:2004vg}
Flaugher B.,  2005, Int.J.Mod.Phys., A20, 3121

\bibitem[\protect\citeauthoryear{{Francis} \& {Peacock}}{{Francis} \&
  {Peacock}}{2010}]{2010MNRAS.406....2F}
{Francis} C.~L.,  {Peacock} J.~A.,  2010, \mnras, 406, 2, \eprint{0909.2494}

\bibitem[\protect\citeauthoryear{{Frith}, {Outram} \& {Shanks}}{{Frith}
  et~al.}{2005}]{2005MNRAS.364..593F}
{Frith} W.~J.,  {Outram} P.~J.,    {Shanks} T.,  2005, \mnras, 364, 593,
  \eprint{astro-ph/0507215}

\bibitem[\protect\citeauthoryear{{Gibelyou} \& {Huterer}}{{Gibelyou} \&
  {Huterer}}{2012}]{2012MNRAS.427.1994G}
{Gibelyou} C.,  {Huterer} D.,  2012, \mnras, 427, 1994, \eprint{1205.6476}

\bibitem[\protect\citeauthoryear{{G{\'o}rski}, {Hivon}, {Banday}, {Wandelt},
  {Hansen}, {Reinecke} \& {Bartelmann}}{{G{\'o}rski}
  et~al.}{2005}]{2005ApJ...622..759G}
{G{\'o}rski} K.~M.,  {Hivon} E.,  {Banday} A.~J.,  {Wandelt} B.~D.,  {Hansen}
  F.~K.,  {Reinecke} M.,    {Bartelmann} M.,  2005, \apj, 622, 759,
  \eprint{astro-ph/0409513}

\bibitem[\protect\citeauthoryear{{Guzzo}}{{Guzzo}}{1997}]{1997NewA....2..517G}
{Guzzo} L.,  1997, \na, 2, 517, \eprint{astro-ph/9711206}

\bibitem[\protect\citeauthoryear{{Hambly} et~al.,}{{Hambly}
  et~al.}{2001}]{2001MNRAS.326.1279H}
{Hambly} N.~C.  et~al., 2001, \mnras, 326, 1279, \eprint{astro-ph/0108286}

\bibitem[\protect\citeauthoryear{{Hartlap}, {Simon} \& {Schneider}}{{Hartlap}
  et~al.}{2007}]{2007A&A...464..399H}
{Hartlap} J.,  {Simon} P.,    {Schneider} P.,  2007, \aap, 464, 399,
  \eprint{astro-ph/0608064}

\bibitem[\protect\citeauthoryear{{Hirata}}{{Hirata}}{2009}]{2009JCAP...09..011H}
{Hirata} C.~M.,  2009, \jcap, 9, 11, \eprint{0907.0703}

\bibitem[\protect\citeauthoryear{{Hoftuft}, {Eriksen}, {Banday}, {G{\'o}rski},
  {Hansen} \& {Lilje}}{{Hoftuft} et~al.}{2009}]{2009ApJ...699..985H}
{Hoftuft} J.,  {Eriksen} H.~K.,  {Banday} A.~J.,  {G{\'o}rski} K.~M.,  {Hansen}
  F.~K.,    {Lilje} P.~B.,  2009, \apj, 699, 985, \eprint{0903.1229}

\bibitem[\protect\citeauthoryear{{Hogg}, {Eisenstein}, {Blanton}, {Bahcall},
  {Brinkmann}, {Gunn} \& {Schneider}}{{Hogg}
  et~al.}{2005}]{2005ApJ...624...54H}
{Hogg} D.~W.,  {Eisenstein} D.~J.,  {Blanton} M.~R.,  {Bahcall} N.~A.,
  {Brinkmann} J.,  {Gunn} J.~E.,    {Schneider} D.~P.,  2005, \apj, 624, 54,
  \eprint{astro-ph/0411197}

\bibitem[\protect\citeauthoryear{{Huchra} et~al.,}{{Huchra}
  et~al.}{2012}]{2012ApJS..199...26H}
{Huchra} J.~P.  et~al., 2012, \apjs, 199, 26, \eprint{1108.0669}

\bibitem[\protect\citeauthoryear{{Itoh}, {Yahata} \& {Takada}}{{Itoh}
  et~al.}{2010}]{2010PhRvD..82d3530I}
{Itoh} Y.,  {Yahata} K.,    {Takada} M.,  2010, \prd, 82, 043530,
  \eprint{0912.1460}

\bibitem[\protect\citeauthoryear{{Jarrett}}{{Jarrett}}{2004}]{2004PASA...21..396J}
{Jarrett} T.,  2004, \pasa, 21, 396, \eprint{astro-ph/0405069}

\bibitem[\protect\citeauthoryear{{Jarrett}, {Chester}, {Cutri}, {Schneider},
  {Skrutskie} \& {Huchra}}{{Jarrett} et~al.}{2000}]{2000AJ....119.2498J}
{Jarrett} T.~H.,  {Chester} T.,  {Cutri} R.,  {Schneider} S.,  {Skrutskie} M.,
    {Huchra} J.~P.,  2000, \aj, 119, 2498, \eprint{astro-ph/0004318}

\bibitem[\protect\citeauthoryear{{Joyce}, {Montuori} \& {Labini}}{{Joyce}
  et~al.}{1999}]{1999ApJ...514L...5J}
{Joyce} M.,  {Montuori} M.,    {Labini} F.~S.,  1999, \apjl, 514, L5,
  \eprint{astro-ph/9901290}

\bibitem[\protect\citeauthoryear{{Joyce}, {Sylos Labini}, {Gabrielli},
  {Montuori} \& {Pietronero}}{{Joyce} et~al.}{2005}]{2005A&A...443...11J}
{Joyce} M.,  {Sylos Labini} F.,  {Gabrielli} A.,  {Montuori} M.,
  {Pietronero} L.,  2005, \aap, 443, 11, \eprint{astro-ph/0501583}

\bibitem[\protect\citeauthoryear{{Keenan}, {Barger}, {Cowie}, {Wang}, {Wold} \&
  {Trouille}}{{Keenan} et~al.}{2012}]{2012ApJ...754..131K}
{Keenan} R.~C.,  {Barger} A.~J.,  {Cowie} L.~L.,  {Wang} W.-H.,  {Wold} I.,
  {Trouille} L.,  2012, \apj, 754, 131, \eprint{1207.1588}

\bibitem[\protect\citeauthoryear{{Keenan}, {Trouille}, {Barger}, {Cowie} \&
  {Wang}}{{Keenan} et~al.}{2010}]{2010ApJS..186...94K}
{Keenan} R.~C.,  {Trouille} L.,  {Barger} A.~J.,  {Cowie} L.~L.,    {Wang}
  W.-H.,  2010, \apjs, 186, 94, \eprint{0912.3090}

\bibitem[\protect\citeauthoryear{{Kitaura}, {Jasche} \& {Metcalf}}{{Kitaura}
  et~al.}{2010}]{2010MNRAS.403..589K}
{Kitaura} F.-S.,  {Jasche} J.,    {Metcalf} R.~B.,  2010, \mnras, 403, 589,
  \eprint{0911.1407}

\bibitem[\protect\citeauthoryear{{Kurokawa}, {Morikawa} \& {Mouri}}{{Kurokawa}
  et~al.}{2001}]{2001A&A...370..358K}
{Kurokawa} T.,  {Morikawa} M.,    {Mouri} H.,  2001, \aap, 370, 358

\bibitem[\protect\citeauthoryear{{Landy} \& {Szalay}}{{Landy} \&
  {Szalay}}{1993}]{1993ApJ...412...64L}
{Landy} S.~D.,  {Szalay} A.~S.,  1993, \apj, 412, 64

\bibitem[\protect\citeauthoryear{{Lewis}, {Challinor} \& {Lasenby}}{{Lewis}
  et~al.}{2000}]{2000ApJ...538..473L}
{Lewis} A.,  {Challinor} A.,    {Lasenby} A.,  2000, \apj, 538, 473,
  \eprint{astro-ph/9911177}

\bibitem[\protect\citeauthoryear{{Limber}}{{Limber}}{1953}]{1953ApJ...117..134L}
{Limber} D.~N.,  1953, \apj, 117, 134

\bibitem[\protect\citeauthoryear{{Maller}, {McIntosh}, {Katz} \&
  {Weinberg}}{{Maller} et~al.}{2003}]{2003ApJ...598L...1M}
{Maller} A.~H.,  {McIntosh} D.~H.,  {Katz} N.,    {Weinberg} M.~D.,  2003,
  \apjl, 598, L1, \eprint{astro-ph/0303592}

\bibitem[\protect\citeauthoryear{{Maller}, {McIntosh}, {Katz} \&
  {Weinberg}}{{Maller} et~al.}{2005}]{2005ApJ...619..147M}
{Maller} A.~H.,  {McIntosh} D.~H.,  {Katz} N.,    {Weinberg} M.~D.,  2005,
  \apj, 619, 147, \eprint{astro-ph/0304005}

\bibitem[\protect\citeauthoryear{Mart\'inez \& Saar}{Mart\'inez \&
  Saar}{2002}]{Martinez:2002}
Mart\'inez V.,  Saar E.,  2002, Statistics of the galaxy distribution.
CRC Press

\bibitem[\protect\citeauthoryear{{Nadathur}}{{Nadathur}}{2013}]{2013MNRAS.434..398N}
{Nadathur} S.,  2013, \mnras, 434, 398, \eprint{1306.1700}

\bibitem[\protect\citeauthoryear{{Pan} \& {Coles}}{{Pan} \&
  {Coles}}{2000}]{2000MNRAS.318L..51P}
{Pan} J.,  {Coles} P.,  2000, \mnras, 318, L51, \eprint{astro-ph/0008240}

\bibitem[\protect\citeauthoryear{{Park}}{{Park}}{2004}]{2004MNRAS.349..313P}
{Park} C.-G.,  2004, \mnras, 349, 313, \eprint{astro-ph/0307469}

\bibitem[\protect\citeauthoryear{Peebles}{Peebles}{1980}]{BookPeeblesLSS}
Peebles P.,  1980, The Large-Scale Structure of the Universe.
Princeton University Press

\bibitem[\protect\citeauthoryear{Peebles}{Peebles}{1993}]{BookPeeblesCosmo}
Peebles P.,  1993, Principles of Physical cosmology.
Princeton University Press

\bibitem[\protect\citeauthoryear{{Planck Collaboration} et~al.,}{{Planck
  Collaboration} et~al.}{2014a}]{2014A&A...571A..16P}
{Planck Collaboration} et~al., 2014a, \aap, 571, A16, \eprint{1303.5076}

\bibitem[\protect\citeauthoryear{{Planck Collaboration} et~al.,}{{Planck
  Collaboration} et~al.}{2014b}]{2014A&A...571A..23P}
{Planck Collaboration} et~al., 2014b, \aap, 571, A23, \eprint{1303.5083}

\bibitem[\protect\citeauthoryear{{Pullen} \& {Hirata}}{{Pullen} \&
  {Hirata}}{2010}]{2010JCAP...05..027P}
{Pullen} A.~R.,  {Hirata} C.~M.,  2010, \jcap, 5, 27, \eprint{1003.0673}

\bibitem[\protect\citeauthoryear{{Quartin} \& {Notari}}{{Quartin} \&
  {Notari}}{2015}]{2015JCAP...01..008Q}
{Quartin} M.,  {Notari} A.,  2015, \jcap, 1, 8, \eprint{1408.5792}

\bibitem[\protect\citeauthoryear{{R{\"a}s{\"a}nen}}{{R{\"a}s{\"a}nen}}{2011}]{2011CQGra..28p4008R}
{R{\"a}s{\"a}nen} S.,  2011, Classical and Quantum Gravity, 28, 164008,
  \eprint{1102.0408}

\bibitem[\protect\citeauthoryear{{Sandage}, {Tammann} \& {Hardy}}{{Sandage}
  et~al.}{1972}]{1972ApJ...172..253S}
{Sandage} A.,  {Tammann} G.~A.,    {Hardy} E.,  1972, \apj, 172, 253

\bibitem[\protect\citeauthoryear{{Sarkar}, {Yadav}, {Pandey} \&
  {Bharadwaj}}{{Sarkar} et~al.}{2009}]{2009MNRAS.399L.128S}
{Sarkar} P.,  {Yadav} J.,  {Pandey} B.,    {Bharadwaj} S.,  2009, \mnras, 399,
  L128, \eprint{0906.3431}

\bibitem[\protect\citeauthoryear{{Schlegel}, {Finkbeiner} \&
  {Davis}}{{Schlegel} et~al.}{1998}]{1998ApJ...500..525S}
{Schlegel} D.~J.,  {Finkbeiner} D.~P.,    {Davis} M.,  1998, \apj, 500, 525,
  \eprint{astro-ph/9710327}

\bibitem[\protect\citeauthoryear{{Scrimgeour} et~al.,}{{Scrimgeour}
  et~al.}{2012}]{2012MNRAS.425..116S}
{Scrimgeour} M.~I.  et~al., 2012, \mnras, 425, 116, \eprint{1205.6812}

\bibitem[\protect\citeauthoryear{{Seshadri}}{{Seshadri}}{2005}]{2005BASI...33....1S}
{Seshadri} T.~R.,  2005, Bulletin of the Astronomical Society of India, 33, 1

\bibitem[\protect\citeauthoryear{{Skrutskie} et~al.,}{{Skrutskie}
  et~al.}{2006}]{2006AJ....131.1163S}
{Skrutskie} M.~F.,  et~al., 2006, \apj, 131, 1163

\bibitem[\protect\citeauthoryear{{Sylos Labini}}{{Sylos
  Labini}}{2011a}]{2011CQGra..28p4003S}
{Sylos Labini} F.,  2011a, Classical and Quantum Gravity, 28, 164003,
  \eprint{1103.5974}

\bibitem[\protect\citeauthoryear{{Sylos Labini}}{{Sylos
  Labini}}{2011b}]{2011EL.....9659001S}
{Sylos Labini} F.,  2011b, EPL (Europhysics Letters), 96, 59001,
  \eprint{1110.4041}

\bibitem[\protect\citeauthoryear{{Sylos Labini}, {Tekhanovich} \&
  {Baryshev}}{{Sylos Labini} et~al.}{2014}]{2014JCAP...07..035S}
{Sylos Labini} F.,  {Tekhanovich} D.,    {Baryshev} Y.~V.,  2014, \jcap, 7, 35,
  \eprint{1406.5899}

\bibitem[\protect\citeauthoryear{{White}, {Tinker} \& {McBride}}{{White}
  et~al.}{2014}]{2014MNRAS.437.2594W}
{White} M.,  {Tinker} J.~L.,    {McBride} C.~K.,  2014, \mnras, 437, 2594,
  \eprint{1309.5532}

\bibitem[\protect\citeauthoryear{{Wright} et~al.,}{{Wright}
  et~al.}{2010}]{2010AJ....140.1868W}
{Wright} E.~L.  et~al., 2010, \aj, 140, 1868, \eprint{1008.0031}

\bibitem[\protect\citeauthoryear{{Xu}, {Cuesta}, {Padmanabhan}, {Eisenstein} \&
  {McBride}}{{Xu} et~al.}{2013}]{2013MNRAS.431.2834X}
{Xu} X.,  {Cuesta} A.~J.,  {Padmanabhan} N.,  {Eisenstein} D.~J.,    {McBride}
  C.~K.,  2013, \mnras, 431, 2834, \eprint{1206.6732}

\bibitem[\protect\citeauthoryear{{Yadav}, {Bagla} \& {Khandai}}{{Yadav}
  et~al.}{2010}]{2010MNRAS.405.2009Y}
{Yadav} J.~K.,  {Bagla} J.~S.,    {Khandai} N.,  2010, \mnras, 405, 2009,
  \eprint{1001.0617}

\bibitem[\protect\citeauthoryear{{Yoon}, {Huterer}, {Gibelyou}, {Kov{\'a}cs} \&
  {Szapudi}}{{Yoon} et~al.}{2014}]{2014MNRAS.445L..60Y}
{Yoon} M.,  {Huterer} D.,  {Gibelyou} C.,  {Kov{\'a}cs} A.,    {Szapudi} I.,
  2014, \mnras, 445, L60, \eprint{1406.1187}

\end{thebibliography}

\appendix
\section{Mock $\Lambda$CDM catalogues}\label{app:mocks_lcdm}
  The possible deviations with respect to statistical homogeneity explored in this
  paper must be evaluated in terms of the statistical uncertainties already allowed
  by the presence of clustering anisotropies in the standard cosmological model.
  Although there exist analytical approximations to calculate these uncertainties
  \citep{2011MNRAS.414..329C,2013MNRAS.431.2834X}, the most reliable method to estimate
  them in the presence of practical complications, such as the complex sky
  coverage of our sample, is to use large ensembles of independent mock catalogues
  reproducing the expected statistical behaviour of our data. These ensembles should
  mimic the properties of the galaxy sample under analysis and cover a similar volume.
  To our knowledge, no public simulations exist with the volume of 2MASS and a sufficiently
  small mass resolution ($M_{\rm halo}\sim10^{11}M_\odot/h$), and even if they did,
  at least $\mathcal{O}(100)$ of them would be needed in order to obtain reliable estimates
  of the uncertainties.
  
  A historically popular alternative method is to generate lognormal
  realizations of the galaxy density field and Poisson-sample them with galaxies
  \citep{2011MNRAS.416.3017B,2011MNRAS.415.2892B}. The lognormal distribution
  has been advocated as a possible model to describe the distribution of the
  non-linear matter density in the Universe \citep{1991MNRAS.248....1C}, and, since
  it is based on locally transforming a Gaussian random field, it can be used
  to generate large numbers of fast independent realizations. Due to its simplicity
  it is easy to guarantee that the mock realizations will reproduce the input
  power spectrum with a very good accuracy \citep{2014MNRAS.437.2594W}, however
  its validity must be carefully assessed on non-linear scales \citep{2010MNRAS.403..589K}.
  
  Although this method has been traditionally used to generate three-dimensional
  realizations, we have adapted it to use only angular information, using an approach
  similar to that of \citet{2010MNRAS.406....2F}. Starting from a map of the angular
  overdensity field in our data $\delta_d(\hat{\bf n})$, computed as described in Section
  \ref{ssec:bias}, the steps used to generate each full realization are:
  \begin{enumerate}
    \item Interpreting $\delta_d(\hat{\bf n})$ as the lognormal counterpart of an underlying
          Gaussian field $\delta_{G,d}(\hat{\bf n})$, we invert the lognormal transformation
          as
          \begin{equation}
           \delta_{G,d}=\log\left[(1+\delta_d)\,\sqrt{1+\sigma_d^2}\right],
          \end{equation}
          where $\sigma_d^2$ is the variance of $\delta_d$.
    \item We compute the power spectrum of the Gaussian density and find its best-fit
          bias as explained in Section \ref{ssec:bias}, finding the values listed in
          the last column of Table \ref{tab:nz_bf}.
    \item We generate a Gaussian realization $\delta_G$ of the best-fit theoretical
          power spectrum using the HEALPix routine {\tt synfast}.
    \item The corresponding log-normal density field is computed as
          \begin{equation}
            1+\delta_{LN}=\exp\left[\delta_G-\sigma_G^2/2\right],
          \end{equation}
          where $\sigma_G^2\equiv\langle\delta_G^2\rangle$.
    \item A discrete number of galaxies are then assigned to each pixel by Poisson-sampling
          the lognormal field with a mean:
          \begin{equation}
            N(\hat{\bf n})=\bar{N}\,\left[1+\delta_{LN}(\hat{\bf n})\right],
          \end{equation}
          where $\bar{N}$ is the mean number of galaxies per pixel in the data. These mock
          galaxies are then distributed at random inside each pixel.
  \end{enumerate}
  We find that this method is able to generate mock catalogues that recover the best-fit
  cosmological power spectrum to excellent precision on linear scales. However care must
  be taken when using them on small scales, where, as noted by \cite{2014MNRAS.437.2594W},
  they may not be able to reproduce the higher-order correlations of the density field.
  The second panel in Figure \ref{fig:maps2} shows the Mollweide projection of one of these
  realizations corresponding to our fiducial sample.

\section{Mock fractal catalogues}\label{app:mocks_fractal}
\begin{figure*}
    \centering
    \includegraphics[height=0.90\textheight]{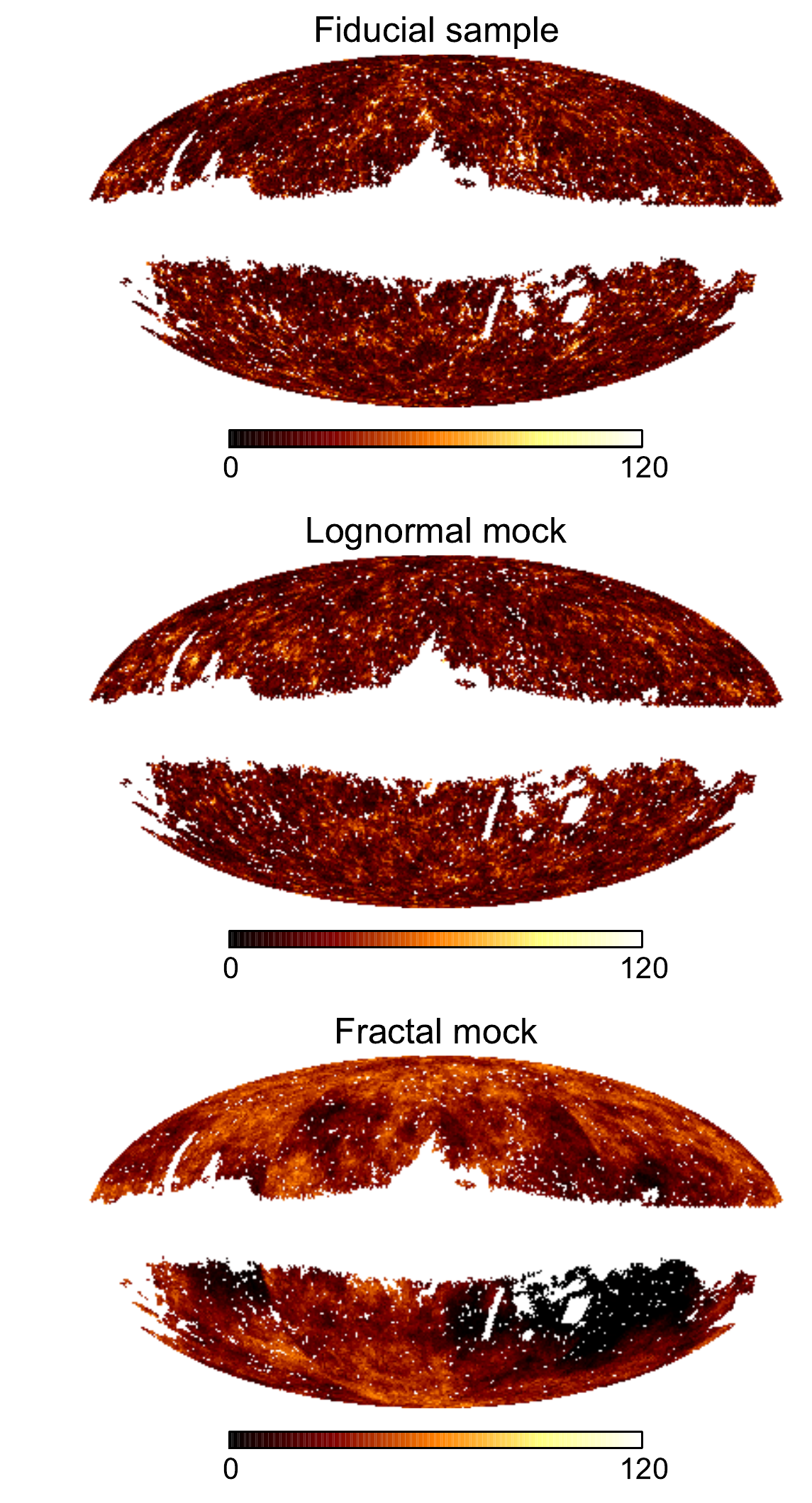}
    \caption{Number density (in ${\rm deg}^{-2}$) of objects in our fiducial sample (upper panel),
             for one of the lognormal realizations described in Appendix \ref{app:mocks_lcdm}
             (middle panel) and for a mock fractal realization with $D=2.75$ (lower panel).}
    \label{fig:maps2}
  \end{figure*}

  The fractal $\beta$-model \citep{1991A&A...246..634C} describes a multiplicative 
  cascading process that is easy to simulate for any fractal dimension $D\leq3$. We
  generated mock realizations of this model with dimensions $D=2.5,\,2.75\,{\rm and}\,2.90$
  using the following method:
  \begin{enumerate}
    \item We divide a cubic box of size $L$, corresponding to twice the maximum comoving
          distance covered by 2MPZ ($\chi_{\rm max}\sim850\,{\rm Mpc}/h$), into 8 sub-boxes
          by dividing each axis in half.
    \item We give each sub-box a probability $p$ of surviving to the next iteration, and
          we select the survivor sub-boxes at random according to that value of $p$. The
          value of $p$ is related to the desired fractal dimension through
          \begin{equation}
            D=3+\log_2p.
          \end{equation}
    \item We repeat the two previous steps on the surviving sub-boxes of the previous
          generation, until we reach the desired resolution.
    \item We place one object at random inside each surviving box in the final set, and
          assign a $K_s-$band luminosity to each of these objects using the luminosity
          function for 2MASS measured by \citet{2014JCAP...10..070A}. This consists of a
          Schechter function with characteristic absolute magnitude $M_*-5\log h\simeq-23.5$
          and power law index $\alpha\simeq-1.02$.
    \item In terms of this absolute magnitude $M_K$, each object is assigned an apparent $K_s$
          magnitude in terms of its distance $d$ to the observer, located at the centre of the
          box:
          \begin{equation}
            K_s=M_K+5\log_{10}\left(\frac{d}{1\,{\rm Mpc}}\right)+25.
          \end{equation}
          Only objects with magnitude $K_s\leq13.9$ are included in the final catalogue.
  \end{enumerate}
  The third panel in Figure \ref{fig:maps2} shows an example of one of these mock realizations
  for $D=2.75$. Note that we do not intend for this model to constitute a viable alternative to
  $\Lambda$CDM, but only to use it as a toy model to verify the validity of the methods applied
  to our galaxy sample. For the same reason, the fact that we use a luminosity function which was
  derived assuming a FRW background should not affect our final results.
  
\end{document}